\documentclass[12pt,preprint]{aastex}
\begin{document}

\title{On the He~II Emission In $\eta$ Carinae and the Origin of Its
  Spectroscopic Events\altaffilmark{1}}
\author{John C.\ Martin\altaffilmark{2}} 
\author{K.\ Davidson\altaffilmark{2}}
\author{Roberta M.\ Humphreys\altaffilmark{2}}
\author{D.\ J.\ Hillier\altaffilmark{3}}
\author{K.\ Ishibashi\altaffilmark{4}}

\altaffiltext{1}{This research is part of the Hubble Space Telescope Treasury Project for Eta Carinae, supported by grants GO-9420 and GO-9973 from the Space Telescope Science Institute (STScI), which is operated by the Association of Universities for Research in Astronomy, Inc., under NASA contract NAS 5-26555. }
\altaffiltext{2}{School of Physics and Astronomy, University of Minnesota, 116 Church Street SE, Minneapolis, MN 55455; martin@etacar.umn.edu}
\altaffiltext{3}{Department of Physics and Astronomy, University of Pittsburgh, 3941 O'Hara Street, Pittsburgh, PA 15260}
\altaffiltext{4}{Center for Space Research, Massachusetts Institute of Technology, 77 Massachusetts Avenue, NE80-6011, Cambridge, MA 02139}

\begin{abstract}
We describe and analyze Hubble Space Telescope (HST) observations
of transient emission near 4680 {\AA} in $\eta$ Car, reported
earlier by \citet{steiner}.  If, as seems probable, this is
\ion{He}{2} $\lambda$4687, then it is a unique clue to $\eta$ Car's 
5.5-year cycle.  According to our analysis, {\it several aspects 
of this feature support a mass-ejection model of the observed 
spectroscopic events,\/} and not an eclipse model.  The \ion{He}{2}
emission appeared in early 2003, grew to a brief maximum during 
the 2003.5 spectroscopic event, and then abruptly disappeared.
It did not appear in any other HST spectra before or after
the event.  The peak brightness was larger than previously
reported, and is difficult to explain even if one allows for
an uncertainty factor of order 3.  The stellar wind must 
provide a temporary larger-than-normal energy supply, and we 
describe a special form of radiative amplification that may 
also be needed.  These characteristics are consistent with
a class of mass-ejection or wind-disturbance scenarios, which
have implications for the physical structure and stability of
$\eta$ Car. 
\end{abstract}

\keywords{binaries: general, line: profiles, stars: individual ($\eta$ Carinae), stars: variables: other, stars: winds, outflows}

\section{Introduction\label{intro}}

\citet{steiner} reported \ion{He}{2}
$\lambda$4687 emission\footnote{In this paper we consistently quote
  vacuum wavelengths.} 
in ground-based spectra of $\eta$ Carinae.  Since \ion{He}{2} represents 
a far higher excitation or ionization state than other lines normally
seen in this object, it may be a valuable clue to $\eta$ Car's  
mysterious 5.5-year cycle and to the structure of strong shock 
fronts in the extraordinary stellar wind (see refs.\ cited below).  
Here we report observations of the same feature with higher spatial 
resolution and other advantages using the Space Telescope Imaging 
Spectrograph (HST/STIS). We find: 

\begin{itemize}
  \item{During the first half of 2003, broad emission appeared 
  between 4675 {\AA} and 4695 {\AA}.  It peaked at the time of the
  mid-2003 ``spectroscopic event,'' with a maximum equivalent width 
  close to 2.4 {\AA}.  \ion{He}{2} $\lambda$4687 is the probable 
  identification.}

  \item{The emission was not spatially resolved from the 
  central star.  Therefore it originated in the dense stellar wind, 
  not in diffuse ejecta at larger radii.}

  \item{This feature was not present in any HST/STIS 
  observation before 2003.0, nor after 2003.5, with a 1-sigma 
  detection limit of about 40 m{\AA} for each occasion and 
  15 m{\AA} for the average.}
    
  \item{Our measurement of the peak brightness considerably 
  exceeds the value reported by Steiner \& Damineli.  The main 
  source of disagreement is explained in Section \ref{sec3} below.}

  \item{Even if one adopts the lower flux estimate, 
  \ion{He}{2}  $\lambda$4687 became far too bright to explain with 
  a conventional model.  The hypothetical companion star's wind
  cannot supply enough energy for the required shock fronts.  The 
  observed emission probably required a mass ejection event on the 
  primary star, and/or some special radiative processes
  described in Section \ref{radexcite} below.}

  \item{Regarding the nature of the spectroscopic event, 
  several aspects of the \ion{He}{2} emission all support a 
  mass-ejection or wind-disturbance type of model.  The observed 
  behavior presents severe difficulties for an ``eclipse'' scenario.}
\end{itemize}

These conclusions are justified in Sections 3--9 below.  
Qualitatively, one expects shock fronts near Eta Carinae to produce
weak \ion{He}{2} emission; but quantitatively {\it the observed flux
  presents a difficult energy-supply problem.\/} Either the deduced
intrinsic strength is in error for reasons unknown, or some
extraordinary process occurred during the 2003 spectroscopic event. 

Since the late 1940's, $\eta$ Car has undergone periodic spectroscopic 
events characterized by the disappearance of high-excitation lines
such as \ion{He}{1} and [\ion{Ne}{3}], both in the stellar wind and in
its nearby slow ejecta. \citet{zanella84} conjectured that these are
essentially mass-ejection episodes.  They recur with a period of 5.5
years \citep{damineli96,whitelock94}, and the emission feature
analyzed in this paper was obviously correlated with the event that
occurred in mid-2003. 

As noted by Steiner and Damineli and discussed below, the \ion{He}{2}
emission problem probably involves X-rays, which vary systematically.
Preceding a spectroscopic event, $\eta$ Car's observable 2--10 keV
{\em hard} X-ray flux rises, becoming increasingly unstable or
chaotic, and then crashes almost to zero \citep{xrays2,xraycurve}.
Softer X-rays, which are crucial for our discussion, may behave quite
differently but cannot be observed because they are absorbed by
intervening material.  The most common explanation for $\eta$ Car's
X-ray production is a colliding-wind binary model.  Some authors
believe that each 5.5-year event involves an eclipse of a hot
secondary star by the primary wind (e.g., see \citet{pc02})
-- instead of a mass ejection event, or, perhaps, in addition to it.
On the other hand, a mass-ejection model, either
binary or single-star, does not require an eclipse
\citep{kd99, kd02}.  The hypothetical companion star has
not been detected, nor have Doppler velocities in the primary wind
provided a valid orbit \citep{davidson00}.  The two alternative
scenarios, eclipse vs.\ mass-ejection, differ in fundamental
significance because the former is simply a result of geometry but the
latter requires an undiagnosed stellar surface instability;  see
remarks in \citet{kd02,kd05c}. 

So far as we know, \citet{gaviola} reported the earliest suspected
detection of \ion{He}{2} $\lambda$4687 in $\eta$ Car.  He listed two
separate emission features near 4680 \mbox{\AA} and 4686 \mbox{\AA}
that were comparable in strength to \ion{He}{1} $\lambda$4714.
According to the notes in his paper, each of them appeared only in one
observation during the interval from April 1944 through March 1951.
Gaviola obtained data during a spectroscopic event in 1948 but
understandably he did not recognize it as such, and his notes do not
state whether that was the time when the emission in question
appeared.  If he observed emission at 4686 \mbox{\AA} on one occasion
in 1948 and then 4680 \mbox{\AA} somewhat later, the wavelength shift
would resemble that which occurred in 2003 (see Section \ref{sec4}).
Unfortunately the published record does not state whether this was the
case, and it is possible that Gaviola's spectra showing this emission
feature did not correspond to a spectroscopic event (see below).

\citet{thackeray} reported that a weak feature occurred near 4687
\mbox{\AA} at some time between April 1951 and June 1952.  Feast 
(2004, {\it private communication}), examining Thackeray's plates,
confirms that such emission may have been present on 1951 June 14
and 1951 July 10.  In itself this evidence is weak, since Thackeray
and Feast both expressed doubt and the observations did not
coincide with an ``event'' in the 5.5-year cycle.  However, we note
that 1951 was an unusual period in $\eta$ Car's recovery from its
Great Eruption.\footnote{
  Beginning about 180 years ago eta Carinae entered a period
  of remarkable variability culminating in its famous ``Great
  Eruption'' from 1837 to 1858 when it became one of the brightest
  stars in the sky. The Great Eruption produced the familiar bipolar
  ``homunculus'' nebula that surrounds the central star.
  See \citet{davidsonhumphreys1997} for many pertinent references.} 
\ion{He}{1} and other moderately high-excitation 
emission had first appeared only a few years earlier 
\citep{feast01,rmh05,gaviola}, and the star brightened rapidly during 
the same era \citep{oc56,dvegg52}.  Thus its wind may have changed from 
one state to another in the years preceding 1953;  see 
\citet{martinconf}, \citet{halpha}, \citet{kd05c}, and refs.\ cited 
therein.  Thus we should not dismiss the possibility that Thackeray's 
plates really did show \ion{He}{2} $\lambda$4687 in 1951.

Since the Hubble Treasury Project for $\eta$ Carinae was intended 
primarily to observe the spectroscopic event in 2003, we obtained STIS 
data repeatedly during that year.   Similar observations had been made 
on a few earlier occasions from 1998 to 2002.  In this paper we discuss 
emission near 4680 \mbox{\AA} in the STIS data;  we also note some pertinent 
ground-based VLT/ESO observations, see \citet{stahl05}.  Often we
refer to the feature in question as ``4680 \mbox{\AA},'' its
approximate peak wavelength at maximum strength, because it is quite
broad and in principle the \ion{He}{2} identification has not been
fully confirmed.  (See Section \ref{aplineid}.)  We also present a
theoretical assessment of the energy budget implied by the line's
strength, which appears almost paradoxical. In general, 
we shall conclude that the observations favor a mass-ejection or 
wind-disturbance scenario, consistent with suggestions 
by \citet{zanella84}, \citet{kd99}, \citet{kd02}, and \citet{nsmith03};  
and that an eclipse does not explain the \ion{He}{2} 
problem.

We describe the observations in Section \ref{sec2}, the observational
constraints on the spatial extent of the emitting region  in Section
\ref{spatial}, identification of the feature in Section
\ref{aplineid}, the emission strength and behavior in Sections 
\ref{sec3} and \ref{sec4}, the theoretical problem in Section
\ref{theorysec} and \ref{radexcite}, and -- most important-- a summary
of implications in Section \ref{conclusions}.  A number of 
technical details and equations are presented in separate
Appendixes. 
\

\section{STIS CCD Data\label{sec2}}

The main spectra in this paper were obtained as part of the $\eta$
Carinae HST Treasury Project \citep{etatp} and were reduced using
a modified version of the Goddard CALSTIS reduction pipeline.
The modified pipeline uses the normal HST bias subtraction, flat 
fielding, and cosmic ray rejection procedures with the addition of
improved pixel interpolation and improved bad/hot pixel removal.
Information regarding these modifications can be found online at our
web site\footnote{\tt http://etacar.umn.edu} and in a forthcoming
publication (Davidson et al., {\it in preparation\/}).  
Each one-dimensional STIS spectrum discussed here is essentially
a 0.1{\arcsec} $\times$ 0.25{\arcsec} spatial sample:  the pixel
size was 0.05{\arcsec}, the slit width was about 2 CCD columns,
and each spectral extraction sampled 5 CCD rows.  Complex details
of our pixel interpolation methods, etc., do not materially affect 
these results.  The CCD column width 
corresponded to a wavelength interval of about 0.28 {\AA} or 
18 km s$^{-1}$.  The instrument's spectral resolution was roughly 
40 km s$^{-1}$ (FWHM), much narrower than the 4680 {\AA} emission 
feature.   We applied an aperture correction to the absolute flux,
based on an observation of the spectrophotometric standard star
Feige 110 with the same slit and extraction techniques.  Absolute
flux measurements, however, are not critical for our results;
the worst uncertainties involve other nearby emission lines
which affected the choice of continuum level relative to the 
overall spectrum.

It is important to note that the HST/STIS resolves the central star
from nearby bright ejecta that heavily contaminate all ground-based
spectra of $\eta$ Car.  Therefore, unlike ground-based observations,
these spectra specifically exclude material outside $r \approx$ 300
AU.   When we refer to ``the star'' or ``$\eta$ Carinae,'' we mean the
central object and its wind, not the diffuse ejecta and Homunculus
nebula. If it is a 5.5-year binary system, then the two stars are not
separated by the HST.  

Experience shows that standard data-reduction software often 
underestimates the measurement errors in high-S/N cases. Therefore 
we estimated our r.m.s.\ noise levels by carefully assessing the 
variations among nearby pixels,  with adjustments for correlations 
caused by interpolation.  Non-statistical ``pattern noise'' caused 
these estimates to be about 30\% larger than we would have gotten 
from only the readout and count-number noise.  Fortunately the 
characteristic size scale of this pattern noise was only a few pixels, 
much narrower than the observed emission feature;  so it affected our 
measurements no worse than statistical noise with the same r.m.s.\ 
amplitude.  In summary, around 4680 {\AA} the overall r.m.s.\ S/N 
ratios in our one-dimensional spectrum extractions ranged from 80 to 
150 for a sample width of 0.28 {\AA} (the CCD column width);  see
Table \ref{ewfluxtab}.  However, the most relevant measurements are
sums or integrations over wavelength intervals of 4 to 19 {\AA}, i.e., 
14 to 70 CCD columns.  Continuum-to-noise ratios formally exceed 
300 for such measurements, but systematic errors presumably dominate 
when such wide samples are taken.  Systematic effects must be assessed 
from other considerations.  In most analyses and figures in this paper, 
we do not average or smooth the spectra;  exceptions will be clearly 
specified.

\section{Mapping the Emission Region\label{spatial}}

Our measurements confirm that the 4680{\AA} feature spatially
coincides with the central star, as previously remarked by \citet{gull}.
To the accuracy of our measurements, the emission is unresolved from
the central star within a radius of the order of 0.03{\arcsec} or
about 70 AU in STIS slits roughly oriented with the presumptive
equatorial or orbital plane (slit angles:  70$\arcdeg$, 62$\arcdeg$,
38$\arcdeg$, and 27$\arcdeg$ measured from north through east).  

\begin{figure}
\figurenum{1}
\label{slitmap}
\includegraphics[scale=.45]{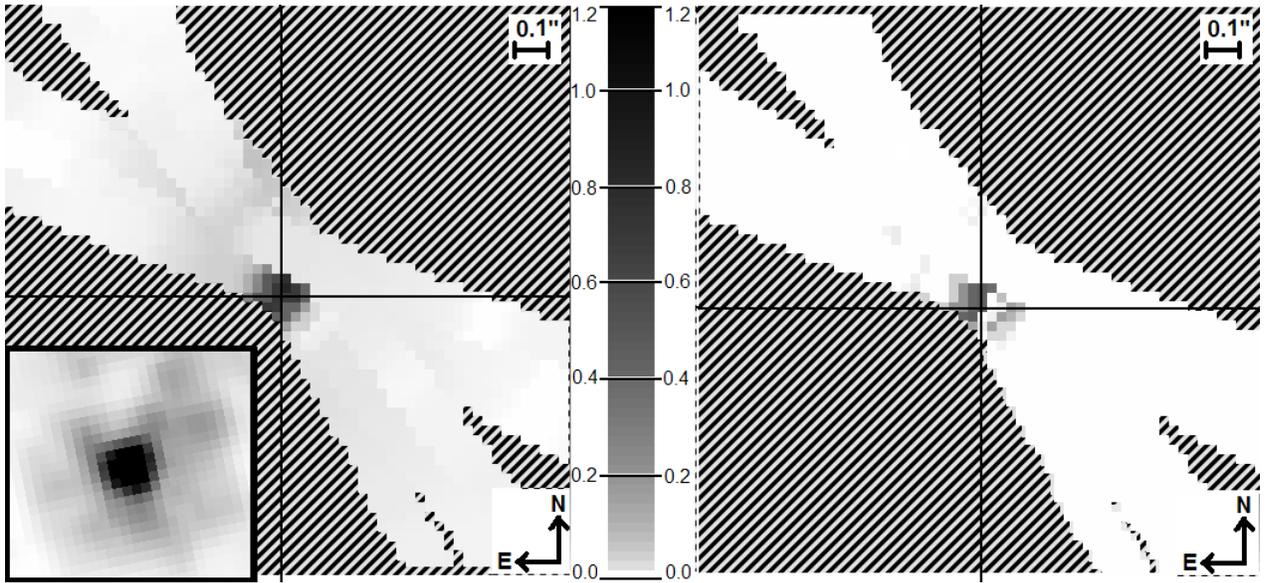}
\caption{Maps of 4680{\AA} emission (left), 4744{\AA} continuum
  (right), and an HST ACS/HRC F330W image at the same scale (insert
  lower left).  The maps are made from seven separate STIS slits
  observed between MJD 52764 (2003.34) and MJD 52813 (2003.47) with
  the relative flux scale in the center.  The diagonal hashed area is
  not covered by any slit and cross-hairs mark the location of the
  central star.  The map is slightly smeared diagonally from lower
  left to upper right because there is no spatial resolution for an
  individual spectrum in the dispersion direction.  The slight
  asymmetry with respect to the position of the central star in the
  map of 4680 \mbox{\AA} emission is a product of the pixel
  interpolation method and is not present in the raw STIS data.}  
\end{figure}

Figure \ref{slitmap} shows that there is no detectable \ion{He}{2}
emission originating in the nearby ejecta.  These data
do not rule out the possibility that emission is extended only
in the southeast direction or at large radii along the polar axis.
However, such a configuration would seem contrived.

\section{Line Identification\label{aplineid}}

Strictly speaking the \ion{He}{2} $\lambda$4687 identification has not 
been confirmed.  Under normal circumstances other \ion{He}{2}
recombination lines should also be present.  However, several factors
conspire to make those lines difficult to detect:
extinction (mostly circumstellar), blends with other spectral features
(esp. hydrogen Balmer lines), and relative weakness of most of the
transitions.  The strongest relevant transitions are listed in Table
\ref{heiitab}.  The detection limit assumes a Gaussian shaped profile
with a Doppler width of $\sim$600 km s$^{-1}$ but the results are
not very sensitive to this assumption.

The strongest accessible transition, $\lambda$1640, does not appear in
the STIS/MAMA/Echelle data \citep{gull}.  This is understandable,
however, since the UV spectrum is complicated by strong overlapping
absorption complexes \citep{hillieretal01,hillier05}.  If the \ion{He}{2}
emission region is, as expected, closely related to the secondary
star's location in its orbit, then the $\lambda$1640 photons may
escape more easily near apastron.  However, our mid-cycle MAMA E140
observation (MJD 51267 = 2000 March 23), shows no sign of
$\lambda$1640.  As explained in Section \ref{wasitthere} we did not
measure any $\lambda$4687 at that time either.

\begin{figure}
\figurenum{2}
\label{compprof}
\includegraphics[scale=.75]{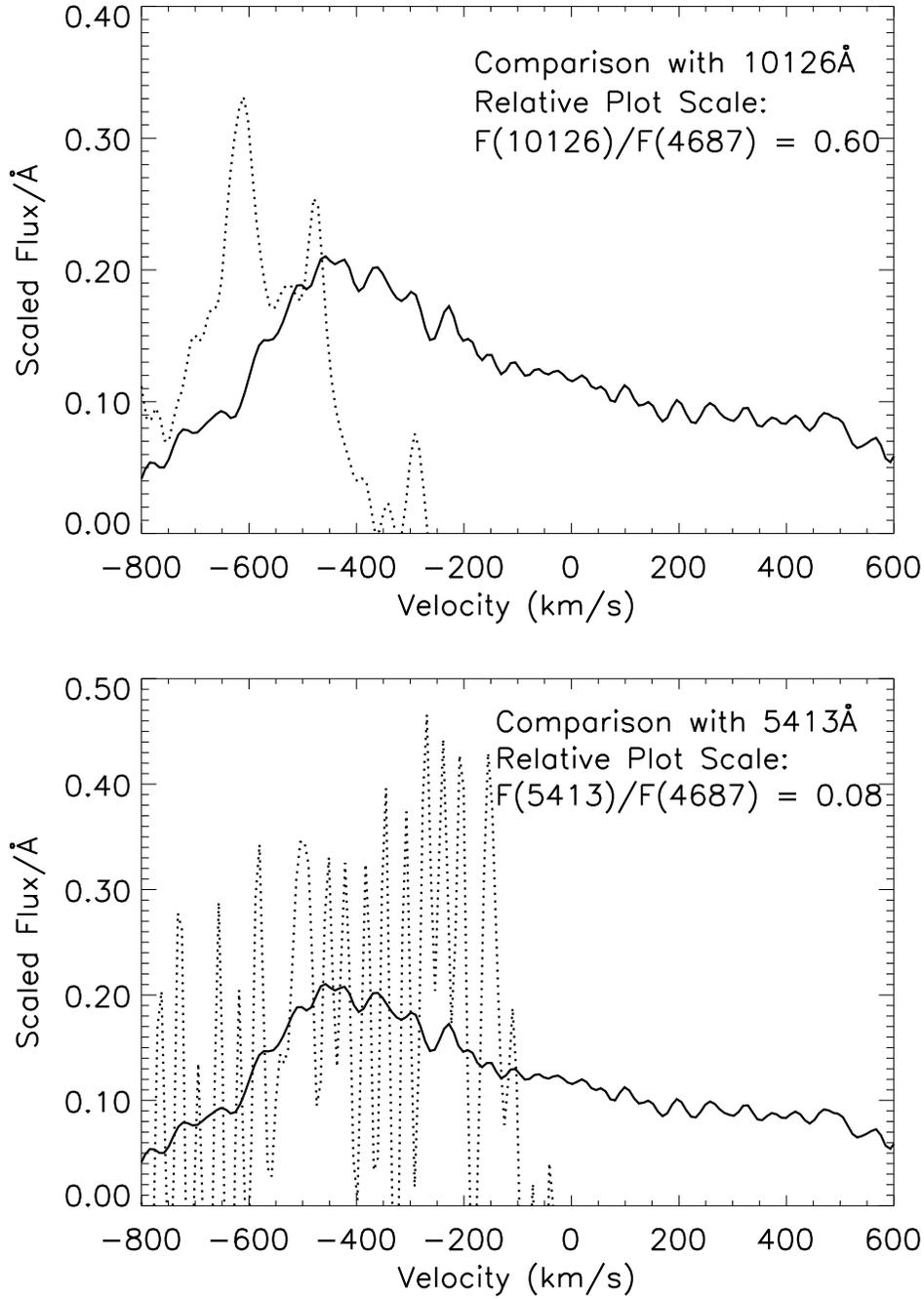}
\caption{A comparison of the observed flux profiles of other \ion{He}{2}
  lines observed on MJD 52813.8 with the profile observed near
  \ion{He}{2} $\lambda$4687 (solid line in both panels).  See text
  for full description.  Top panel:  a comparison with the observed
  flux near \ion{He}{2} $\lambda$10126 (dotted line).  Bottom panel:
  a comparison with the observed flux near \ion{He}{2} $\lambda$5413
  (dotted line).}
\end{figure}

The two other \ion{He}{2} lines most favorable for detection are
$\lambda$5413 and $\lambda$10126.  In Figure \ref{compprof} we have
plotted for each of these as the difference (flux at time of 4680{\AA}
maximum) -- (average of data at times when 4680{\AA} was not
detected).  These differential fluxes are scaled with respect their
respective transition strengths (Table \ref{heiitab}) and extinction
factors \citep{fitz,ccm} relative to $\lambda$4687.

$\lambda$10126 is adversely affected by higher detector noise and
uncertainty in the flux calibration because this wavelength is near
the physical edge of the CCD and the red limit of STIS sensitivity.
There is a significant increase in flux near $\lambda$10126 on MJD
52813.8, but the velocity profile is notably different from
4680{\AA}.  $\lambda$5413 appears to have a similar profile to
4680{\AA} but the noise level negates the significance of the match.
We conclude that we have not detected any additional \ion{He}{2} lines
in our data but the limits are not strong enough to contradict the
$\lambda$4687 identification.  

Under these circumstances it is prudent to examine possible alternative
identifications.  A pair of \ion{N}{3} lines near 4681{\AA} might
contribute to the feature.  However, they are high excitation lines
and our own inspection of the spectrum agrees with \citet{thackeray}
that the species \ion{N}{3} is ``doubtfully present.''  A cluster of
\ion{Ne}{1} lines from 4680\mbox{\AA} to 4683\mbox{\AA} might also be
responsible for the observed emission, but there is no reason to
expect such emission in the stellar wind. Unidentified lines exist
near 4679.21 {\AA} in the Orion Nebula \citep{m42spect} and 4681.38
{\AA} in RR Tel \citep{rrtel};   but it is obviously difficult to
relate the $\eta$ Car feature to another unidentified line in an
object with appreciably different physical conditions. 

\ion{He}{2} $\lambda$4687 seems the ``most natural'' identification 
because it plays a well-known, moderately unusual role in nebular 
astrophysics.  Helium is abundant in $\eta$ Car, and the He$^+$ 
3--4 transition is simple in most respects but has special characteristics 
employed in Section \ref{radexcite} of this paper.  Therefore this
identification is the best working hypothesis.


\section{Flux and Equivalent Width of the Feature\label{sec3}}

\begin{figure}
\figurenum{3}
\label{continuumfig}
\includegraphics[scale=0.75]{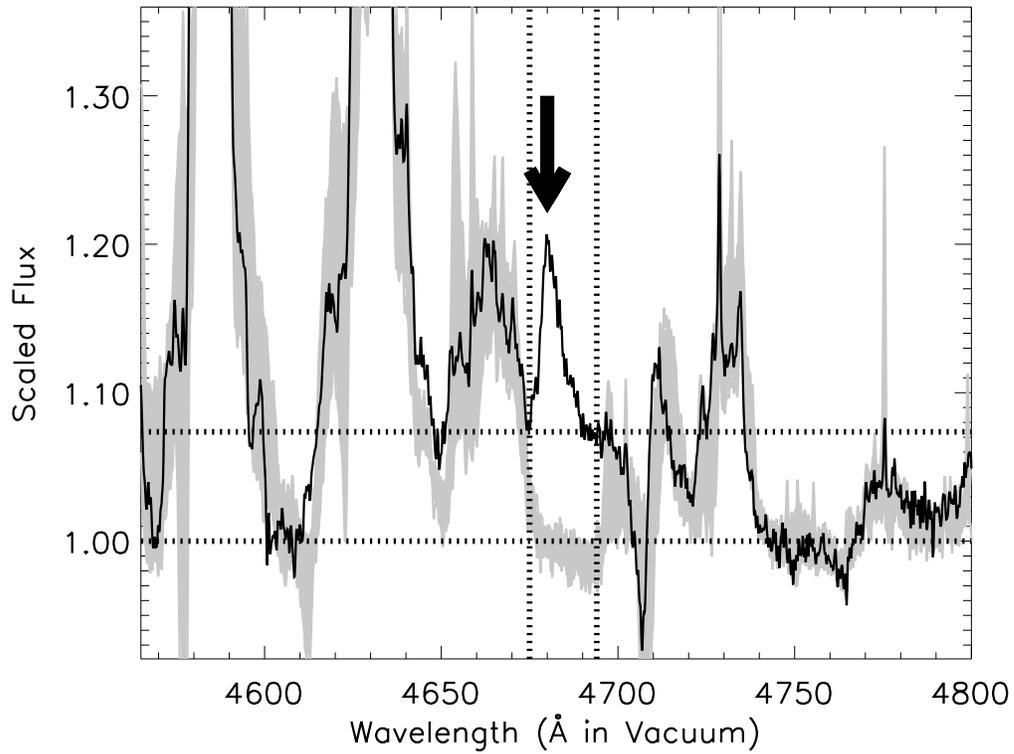}
\caption{A demonstration of the effects of different continuum
  levels (horizontal dotted lines) on the flux measured within the
  integration limits (vertical dotted lines).  The spectrum on MJD 52813.8 
  (solid line) is plotted with the scaled flux range observed when the 
  feature is not present (gray envelope).  Note that we use the 
  the lower continuum level estimated at 4744{\AA}. It is consistent with 
  the continuum near 4605{\AA} as well as the 4680{\AA} flux 
  observed when the feature is not present.}
\end{figure}

Figure \ref{continuumfig} shows that the underlying continuum level
can be a major source of systematic error.  Relatively strong 
features bracket the 4680 {\AA} emission: \ion{Fe}{2} around 
4660 {\AA} to the blue and \ion{He}{1} $\lambda$4714 to the red.
They make it difficult to set the local continuum level when 
the 4680 {\AA} feature becomes strong and broad.  Therefore we 
measured the continuum flux in a separate, nearby wavelength 
interval 4742.5--4746.5 {\AA}, which is devoid of substantial 
features.  This level always agreed well with another likely continuum 
sample near 4605 {\AA}, it matched the flux near 4680 {\AA} 
on every occasion when the emission feature was absent, and 
in general the data give no hint of a strong continuum slope 
or related error (see the lower horizontal line in Fig.\ 
\ref{continuumfig}).   Thus we adopt it as the best available 
measure of the underlying continuum around 4680 {\AA}.  

\begin{figure}
\figurenum{4}
\label{evol4686b}
\includegraphics[scale=0.75]{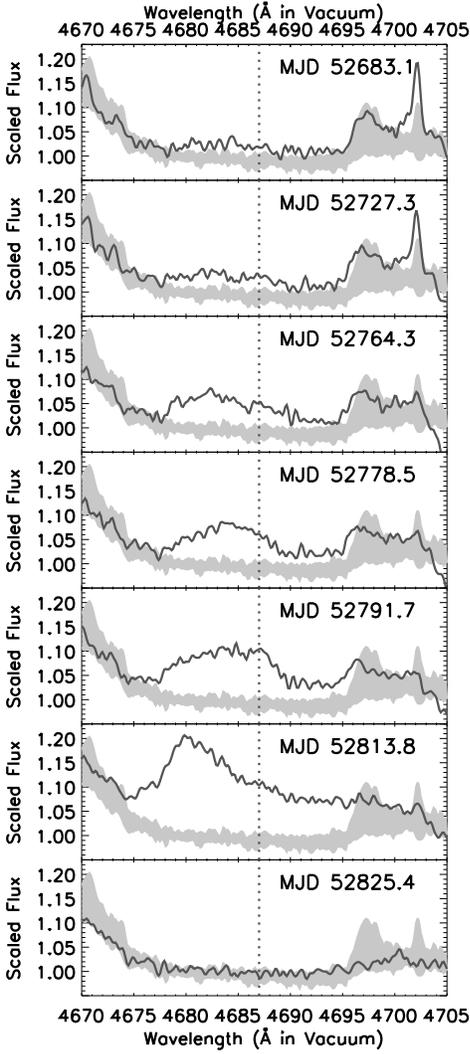}
\caption{A time sequence showing the change in the 4680\mbox{\AA}
  emission feature.  The solid line is the scaled flux on that MJD
  (continuum flux = 1).  For comparison, the gray envelope is the
  range of scaled fluxes observed when the 4680\mbox{\AA} feature was
  not present.  These tracings have not been smoothed.} 
\end{figure}

The adopted continuum levels, continuum-subtracted emission fluxes of
the feature, and corresponding equivalent widths are listed 
in Table \ref{ewfluxtab}.  Our integration range (4675.0--4694.0 {\AA})
omits the extreme wings of the feature in order to exclude
\ion{He}{1} $\lambda$4714 P-Cygni absorption and [\ion{Fe}{3}] 
$\lambda$4702 emission which varied during the spectroscopic event.  
Therefore some of our measurements probably underestimate the flux 
in the 4680 {\AA} feature to a small extent that must be judged from
Figs.\ \ref{continuumfig} and \ref{evol4686b}.  
The $S/N$ column in Table \ref{ewfluxtab} is well defined by the standard
deviation in the continuum, corrected for pixel-to-pixel correlations
caused by interpolation.  The uncertainty
estimates for the line flux and equivalent width are in turn based on the 
$S/N$ and the r.m.s. scatter of the measurements made before 2003 when the 
feature was not present.  These uncertainty estimates omit some 
sources of systematic error that are common to all spectroscopic data, i.e. 
irregularities in the continuum slope, possible weak features everywhere 
in the spectrum, flux calibration errors, variation in slit throughput, 
etc.

The maximum equivalent width that we measured, 2.4 {\AA} at MJD 52813.8,
is more than twice as large as the highest value quoted by
\citet{steiner}.  This disagreement results mainly
from the placement of the the underlying continuum, relative to
the total flux.  The continuum level
adopted by those authors can be deduced from their Fig.\ 1 and their
net equivalent width for the  \ion{He}{2} emission.  At the time when
the $\lambda$4687 feature was brightest, they appear to have used a
continuum near the {\it upper\/} horizontal line in our Fig.\
\ref{continuumfig}; which of course led to a smaller net equivalent
width.\footnote{
  Steiner \& Damineli determined their continuum using a third order
  polynomial, which is affected by line blending and noise in their
  flat fields. They remark that these caused an underestimate of the
  net flux and FWHM of the feature. 
}  
If one views only the wavelength interval 4650--4720 {\AA}
shown in their Fig.\ 1, then the broad wings of the emission feature
are indistinguishable from local continuum; compare MJD 52813 in our
Fig.\ \ref{evol4686b}.  At most other (non-maximum) times, Steiner and
Damineli's continuum levels appear to have been close to ours. 

For $\eta$ Car the worst uncertainties are systematic rather than 
statistical.  Based on the internal consistency of Fig.\
\ref{continuumfig}, and on the absence of serious discrepancies in the
STIS data around 4680 {\AA}, we conclude that the pseudo-sigma
uncertainty in our adopted continuum level {\it at the time of the
  event\/} was of the order of 1\% and no worse than 2\%.  If we
include this in addition to the statistical uncertainties listed in
Table \ref{ewfluxtab}, then, informally and somewhat pessimistically,
the maximum equivalent width was 2.4 $\pm$ 0.4 {\AA}.  Given the
crowded nature of $\eta$ Car's spectrum, there is no evident way to
improve this result.

Some other details are worth noting.  We have verified the absolute
flux calibration in our STIS data to a few percent, and the 4680 {\AA}
feature occurred in an optimal position near the physical center of
the CCD.  In the wavelength range of interest the instrumental
p.s.f. was well focused and the slit-throughput correction was flat.
These factors make us reasonably confident that no unexpectedly large
systematic errors lurk in the data. 

Based on reasoning presented in Appendix \ref{ap_lum}, the peak
intrinsic \ion{He}{2} $\lambda$4687 luminosity was of the order of
10$^{36}$ ergs s$^{-1}$.  To forestall later misunderstandings, we
note that the energy supply problem described in Section
\ref{theorysec} below is {\it not\/} merely a result of the equivalent
width disagreement discussed above, our value vs.\ that of Steiner
and Damineli.  The main conclusions of Section \ref{theorysec} remain
valid even if one adopts their lower estimate.

\section{Temporal Evolution\label{sec4}}

\subsection{Changes in Line Profile}
The first significant detection of the 4680\mbox{\AA} feature in the
STIS data was on MJD 52683.1 (2003 February 12), about 140 days before
the mid-2003 event.  It increased in strength until it reached a
strong maximum on MJD 52813.8 (2003 June 22), at the time of the
event.  During the period of growth its profile undulated
significantly (Figure \ref{evol4686b}).  In general, its width was
roughly 600 km s$^{-1}$ and most of the flux was blueward of the \ion{He}{2}
$\lambda$4687 rest wavelength.  The main profile was usually either
roughly symmetric (MJD 52778.5) or slopped toward the blue (MJD
52791.7).  At maximum, however, a pronounced peak appeared at $-450$
km s$^{-1}$ and the profile slopped toward the red (MJD 52791.7).
\citet{steiner} report similar activity in their observations with
better temporal sampling.  Similar velocity shifts may explain why
\citet{gaviola} reported two separate emission features, if they
appeared in two distinct observations (see Section \ref{intro}). 

\subsection{The Sudden Disappearance\label{gonequick}}
After reaching a strong maximum, the feature abruptly disappeared 
between MJD 52813.8 and 52825.4 (2003 June 22 and July 5), an 
interval of only 12 days.  In fact the disappearance timescale was 
even shorter than that, since Steiner \& Damineli  observed strong 
emission several days after MJD 52813.8. Their data indicate 
a decay time of only about 5 days.   The feature was not detectable
in any of the subsequent STIS observations through MJD 53071.2
(2004 March 6, Fig.\ \ref{nofeature}).   

The VLT/UVES spectra \citep{stahl05} reveal an interesting
wrinkle in the disappearance of the 4680{\AA} feature.  A location 
called FOS4 in the Homunculus Nebula gives a reflected pole-on view 
of the  stellar wind, with a light-travel delay time of about 20 days 
relative to our direct line of sight \citep{meaburn87,kd01,nsmith03}.
Stahl et al.\ show that the 4680{\AA} emission reflected at FOS4
disappeared somewhat earlier (when corrected for light travel time)
and perhaps also more gradually than in the direct view.  This is
difficult to gauge, however, because the VLT/UVES observations were
infrequent compared to the duration of the event.  A comparison with
the STIS data (whose temporal sampling was even more sparse) suggests
that the phenomenon depends on viewing angle; see Figure
\ref{uvesvsstis2}.  These differences merit detailed study because any
serious theoretical model should account for them. 

\subsection{Before and After The Event\label{wasitthere}}
\begin{figure}
\figurenum{5}
\label{nofeature}
\includegraphics[scale=0.4]{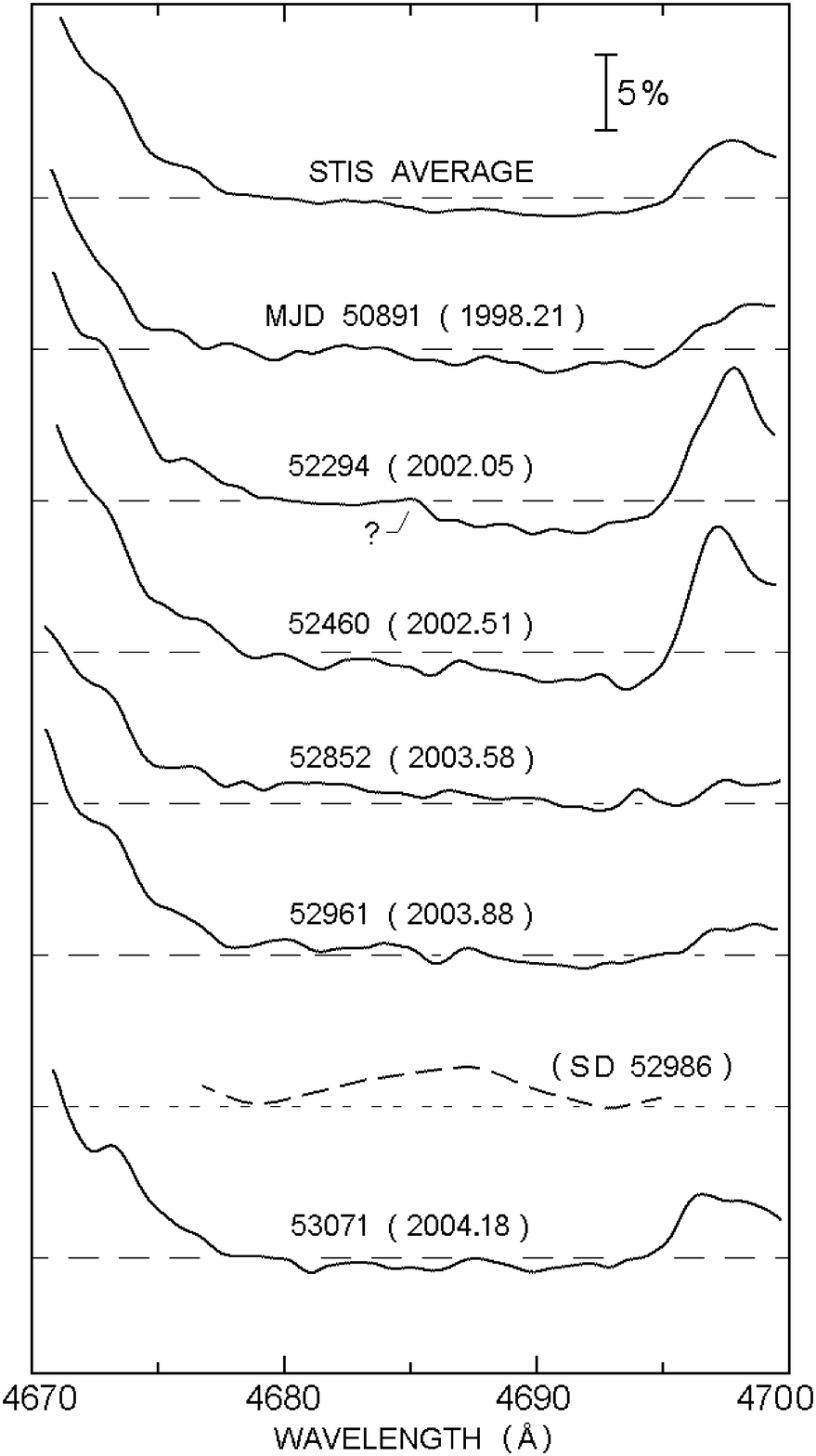}
\caption{A sampling of STIS/CCD spectra from times when
  4680 \mbox{\AA} was not present.  These were smoothed with
 $\Delta \lambda \approx$ 0.5\mbox{\AA}.  At the top is the average scaled
  flux of these data.  Second from the bottom is the spectrum
  reported by \citet{steiner} on MJD 52986 when we detect no trace of
  the feature.  The dashed horizontal lines
  denote the continuum as determined by the average flux between
  4742.5 \mbox{\AA} and 4746.5 \mbox{\AA}.}  
\end{figure}

\begin{figure}
\figurenum{6}
\label{uvesvsstis2}
\includegraphics[scale=0.75]{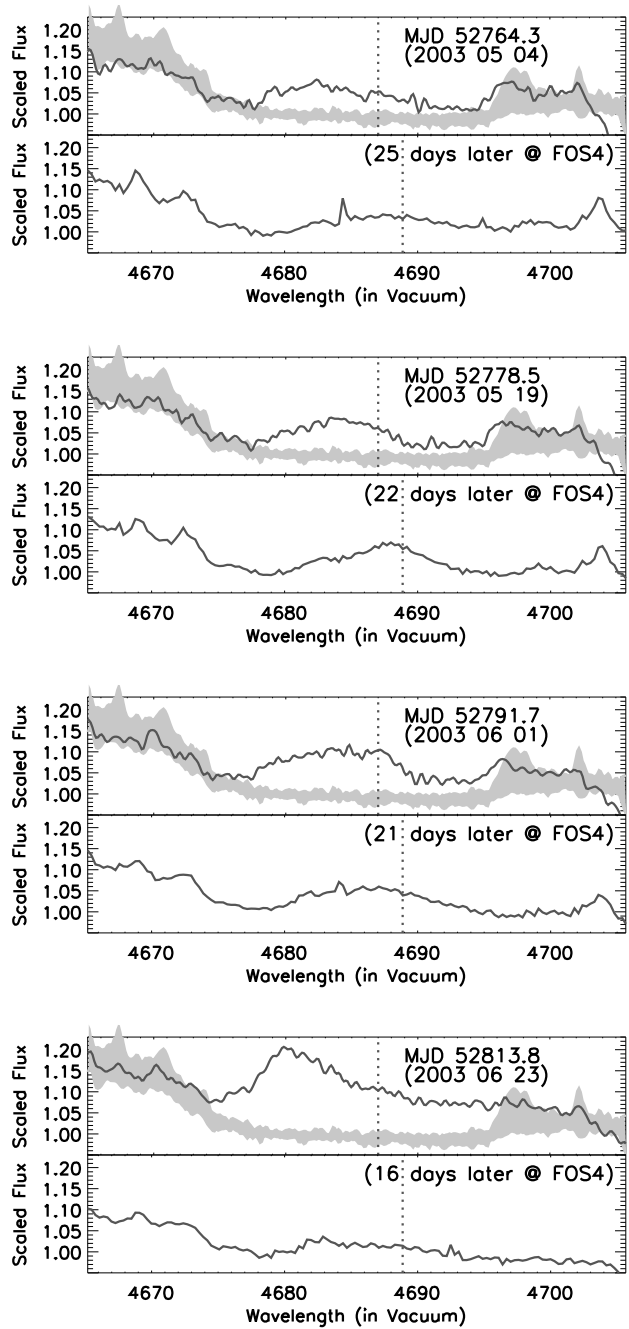}
\caption{A comparison of the STIS spectrum (top panel of each set) of
  the central star with a corresponding VLT/UVES spectrum at a
  location in the Homunculus labeled FOS 4 (bottom
  panel of each set), with the delay noted (picked to be comparable to the
  estimated 20 day difference introduced by light travel time).  Adjacent
  pixels in the VLT/UVES data have been combined so that data is plotted
  with the same spectral resolution as the STIS data ({$\Delta \lambda
  \approx$} 0.3 \mbox{\AA}).  The gray envelope plotted with the STIS
  data is the same as in Figure \ref{evol4686b}.  The vertical dotted
  line marks the rest wavelength of \ion{He}{2} $\lambda$4687 in the
  STIS data and is shifted in the VLT/UVES spectrum to the velocity of the
  reflecting ejecta.} 
\end{figure}

The STIS spectra before 2003.0 and after 2003.50  (MJD before 52640 and
after 52820) all resemble MJD 52825.4, shown at
the bottom of Fig.\ \ref{evol4686b}.  Fig.\ \ref{nofeature} 
illustrates the situation fairly well.  Table \ref{ewfluxtab} shows no
detections at the times in question, and below we obtain stricter
detection limits.  In Table \ref{ewfluxtab}, small  positive and
negative values of net ``emission flux'' at the times  of
non-detection are largely due to changes in the wing of the
\ion{Fe}{2} complex at 4660\mbox{\AA}, which creeps in on the  blue
side of our integration range (Fig.\ \ref{continuumfig}).  Those small
fluctuations mirror the general behavior of broad \ion{Fe}{2} features
in $\eta$ Car's spectrum.

The last column of Table \ref{ewfluxtab} offers independent detection
limits in each observation without any assumptions about the emission
profile.  However, it is possible to obtain stricter limits by
assuming an emission profile with the characteristics specified by
Steiner \& Damineli (see below):  FWHM
$\sim$ 550 km s$^{-1}$ and a central Doppler velocity in the range
from -250 to +150 km s$^{-1}$.  For each STIS observation we performed
a conventional least-squares fit to the function 
$f({\lambda})  =  A  +  B {\lambda}  +  C {\phi}({\lambda})$
over the 57 pixels within the 4677--4693~{\AA} range where
${\phi}({\lambda}$) is the assumed profile.  (We assumed a Gaussian
shape but this has little effect on the results.)  In one respect we
deviated from the proper statistical testing procedure.  Instead of
solving for the best fit central Doppler velocity we choose the value
that resulted in the largest $C$ in the least squares fit.  This
overestimates the suspected emission but for the present
purpose it provides a conservative upper limit.  The uncertainty in $C$
is assessed by a numerical Monte Carlo technique.

The results are presented in Table \ref{tableb1}.  The $S/N$ ratios
and statistical errors are critical here, and we estimated them in two
ways.  One method employed r.m.s.\ differences between wavelength
samples separated by 2, 3, 4, and 5 CCD columns, while the other used
differences between the data and the least-squares fits.  Both methods
agreed well (their average is quoted in Table \ref{tableb1});
moreover, the r.m.s.\ scatter in equivalent width turns  out to be
very close to the expected value based on our error estimates.
Therefore the $S/N$ ratios and uncertainties listed in Table 3 are
quite robust.

As a check, we included one date (MJD 52683) when the feature was
present to verify that our method could detect weak emission.  At
first sight, the equivalent width found for that date seems to
conflict with the value in Table \ref{ewfluxtab}.  However, a rather
large uncertainty is stipulated in Table \ref{ewfluxtab}
and at that time the feature appeared significantly broader than the
assumed profile.  We emphasize that the values in Tables
\ref{ewfluxtab} and \ref{tableb1} were obtained by two very
different, nearly independent methods. 

Aside from the 2003.12 (MJD 52683) observation, Table \ref{tableb1}
shows no evidence for \ion{He}{2} emission before 2003 and after
2003.5.  Each observation individually allows an equivalent width of
40~m{\AA} at the one sigma level; but combining the eleven independent
observations we obtain an average of nearly zero with a formal error
of $\pm$11 m{\AA}.\footnote{
  The last line in Table \ref{tableb1}, $-11 \pm 14$ m{\AA}, refers to
  a pixel-by-pixel sum of the relevant spectra.  It differs from the
  average of the eleven individual equivalent width measurements
  because the individual least-squares fits did not all have the same
  peak velocity.  This distinction is obviously academic, since both
  types of average give null results with $\sigma < 15$ m{\AA}.}

\citet{steiner} reported that the 4680{\AA} emission line 
maintained an equivalent width of 50 to 150 m{\AA} throughout most 
of the spectroscopic cycle.  However, this assertion is not supported
by the STIS data, nor by the VLT/UVES observations described by
\citet{stahl05}.  A specific comparison with Steiner \&
Damineli's data appears near the bottom of Fig.\ \ref{nofeature}.  The
tracing labeled ``SD 52986'' is copied from phase 1.083 in their Fig.\
1.  A feature with that strength would be obvious in the STIS data
even without a formal analysis, e.g., compare our data for MJD 52961
and 53071.  Their measurements could have been affected by some
systematic offset in their continuum (see Section \ref{sec3}).  Or it
is possible that they measured a contribution from any of three
nebular lines near that wavelength which could not have been
resolved from the central star in their data: \ion{N}{1}
$\lambda$4679, \ion{Cr}{2} $\lambda$4680, and [\ion{Fe}{2}]
$\lambda$4688 \citep{zethsonphd}.  They may have observed emission
from an extended region ($r \gtrsim 0.4{\arcsec}$), but if so that
requires an additional emission process, quite different from those
which they considered and we entertain that below.  

\subsection{Timing Relative to Other Phenomena\label{compothers}}

\begin{figure}
\figurenum{7}
\label{comptiming}
\includegraphics[scale=1.]{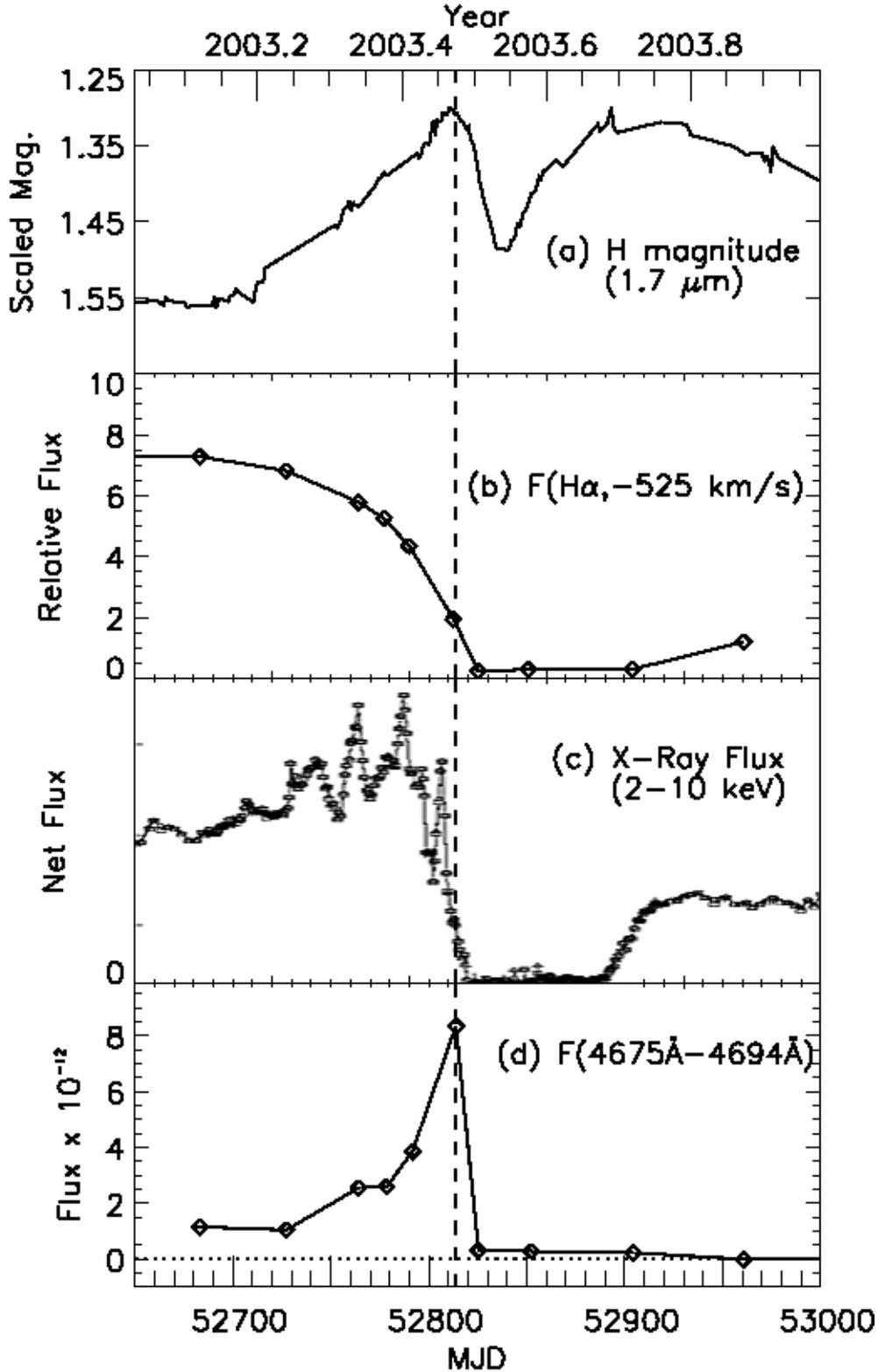}
\caption{A comparison of the temporal evolution of the 4680\mbox{\AA}
  emission feature strength (d) relative to:  the near infrared H
  magnitude brightness \citep{whitelock} (a), the strength of
  the H$\alpha$ P-Cygni absorption \citep{halpha} (b), and the
  2--10 keV X-ray flux \citep{xraycurve} (c).}
\end{figure}

As Figure \ref{comptiming} shows, the development of the 
4680 {\AA} feature coincided with other observed changes in the 
stellar wind.   As its brightness grew during April and May 
of 2003 (MJD $\sim$ 52740 to 52300), so did the 
near-infrared continuum \citep{whitelock} and the depth of the
H$\alpha$ P Cyg absorption \citep{halpha}.   Meanwhile the
2-10 keV X-rays reached their maximum \citep{xraycurve}.
Here we note a significant fact that previous authors have neglected:
{\it The spectroscopic event, including the 4680 {\AA} emission, culminated 
nearly a month after the apparent X-ray maximum.\/}  By the last week
of June 2003, when every major UV-to-IR indicator reached a climax, 
the observed 2-10 keV flux had already declined to a small fraction of 
its peak value.   The dramatic final rise of the 4680 {\AA} feature 
was {\it anti-\/}correlated with the observed hard X-ray flux 
(Fig.\ \ref{comptiming}).\footnote{
   The STIS observation on MJD 52791.7 nearly coincided with the hard 
   X-ray peak.  Therefore, even if the 4680 {\AA} maximum occurred before
   the next STIS observation on MJD 52813.2, the hard X-rays had begun
   their decline.   \citet{steiner} indicate that the peak 4680 {\AA} 
   brightness probably occurred several days {\it after\/} MJD 52813.2. 
   Note that Figure \ref{comptiming} is plotted in terms of MJD rather
   than ``phase'' in the 5.5 year cycle.  This avoids confusion over a
   largely arbitrary zero-point which is not the same in every
   discussion.}
We shall discuss the implications in Section \ref{conclusions}.

\section{The Energy-budget Problem\label{theorysec}}

\subsection{Generalities}

\ion{He}{2} $\lambda$4687 emission is normally a recombination line 
in a He$^{++}$ region, arising in decays from the He$^+$ $n = 4$ 
level to $n = 3$. Eta Car, however, cannot produce sufficient He$^{++}$ 
in the usual way;   the extremely hot stellar continuum required for 
that purpose would also create a bright high-excitation photoionized 
region in the inner Homunculus nebula, contrary to observations.  
Therefore, as \citet{steiner} noted, the He$^{++}$ probably occurs 
near shock fronts that produce ionizing soft X-rays and extreme UV
photons.   Here we review the problem; the $\lambda$4687 brightness 
seen in June 2003 turns out to be more difficult to explain than 
those authors indicated.   A mass ejection event or at least a major 
inner-wind disturbance appears necessary.  Where no specific reference 
is cited here for a parameter, see \citet{osterbrock} for ionic data or 
\citet{davidsonhumphreys1997} and \citet{hillieretal01} for $\eta$ Car.

\begin{figure}
\figurenum{8}
\label{ionzones}
\includegraphics[scale=.5]{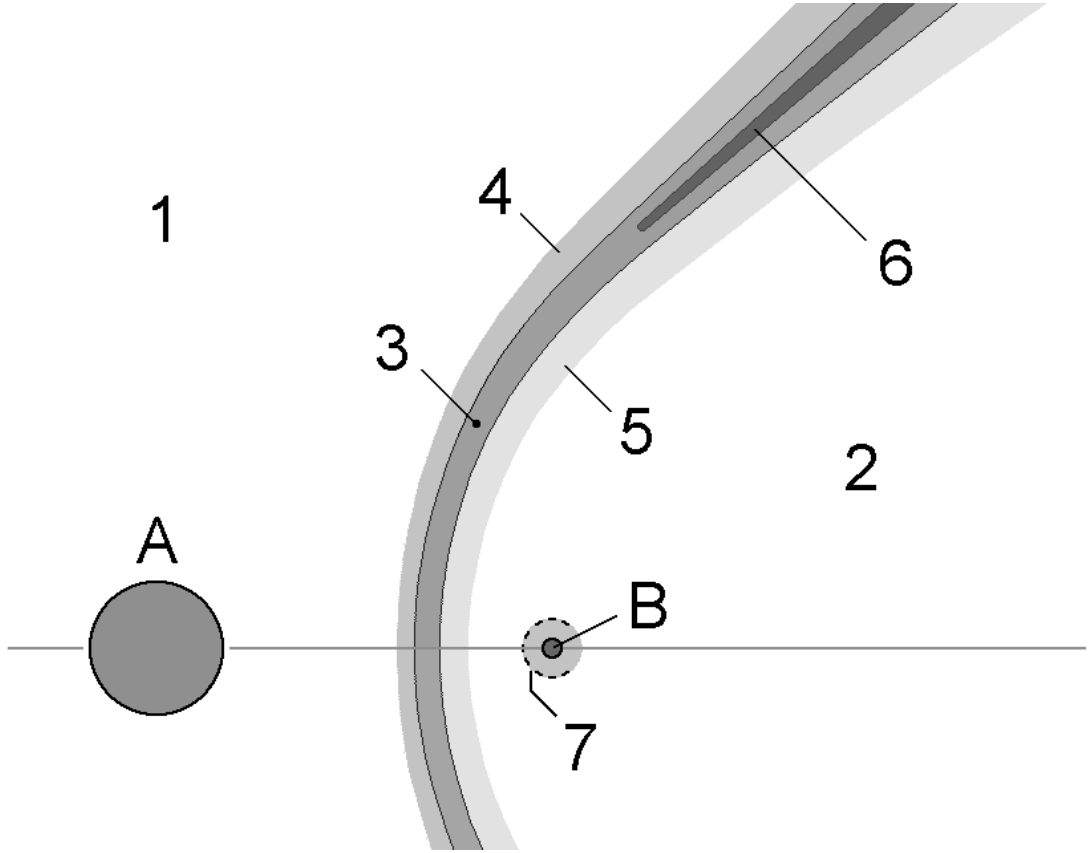}
\caption{Schematic classification of ionization zones that may exist in
  the wind-wind interface between the primary (A) and hypothetical
  secondary (B) stars. See text for a full explanation.  In reality
  the shock fronts and regions do not have simple geometries
  and they certainly vary with time.}
\end{figure}

To put the question in a more definite context, here are three
alternative scenarios for $\eta$ Car's spectroscopic events: 

  \begin{enumerate}
  \item  The shocks may occur at a wind-wind interface in a binary
  system, as most authors have supposed for the observable hard 
  X-rays.  In this case we can classify several distinct emission 
  zones sketched in Fig.\ \ref{ionzones}.  Regions 1 and 2 are 
  undisturbed parts of the two stellar winds, with speeds around 
  500 km s$^{-1}$ and  3000 km s$^{-1}$ respectively.   
  Region 3 contains shocked gas with $T > 10^6$ K.  
  In regions 4 and 5 near the shock surfaces, helium 
  is photoionized by soft X-ray photons from region 3.  Region 6 
  consists of shocked gas that has cooled below $10^5$ K 
  as it flows outward within region 3;  densities are very 
  high there because the local pressure is comparable to the nearby 
  ram pressure of each wind.  (Depending on several parameters, 
  region 6 may be farther from the star than our sketch indicates.)
  In reality each of these ``regions'' probably consists of numerous 
  unstable corrugations or separated condensations (see figures in
  \citet{xrays92} and \citet{pc02}), but the our 
  classification of zones seems broadly valid. There is no 
  reason to expect detectable \ion{He}{2} $\lambda$4687  emission 
  in regions 1 or 2; region 3 is too hot for efficient $\lambda$4687
  emission (see below);  and region 5 has a much smaller density 
  than region 4.  Evidently, then, regions 4 and 6 harbor the best 
  conditions for \ion{He}{2} emission.  Finally, Region 7 in the 
  figure is the acceleration zone of the secondary wind, where 
  \citet{steiner} proposed that the \ion{He}{2} emission originates.
  That idea seems unlikely for reasons that will become evident 
  below.  
  \item   A spectroscopic event of $\eta$ Car may be a stellar 
  mass-ejection phenomenon, possibly including one or more shock 
  fronts moving outward from the primary star 
  \citep{zanella84,kd99,kd02,kd05c}.   In some versions of this story,
  the secondary wind shown in Fig.\ \ref{ionzones} may be 
  unnecessary and the hypothetical companion star might not exist;  
  but in any case an ejection event would produce a relatively large 
  temporary energy supply.   As we emphasize later, this can be
  useful for the \ion{He}{2} $\lambda$4687 emission.  The 
  term ``ejection event''  is rather elastic in this context and 
  may refer to a disturbance in the wind or a temporary 
  alteration of its latitude structure \citep{nsmith03}.   
  \item   Conceptually, at least, it is easy to combine ideas (1) 
  and (2) if the companion star triggers an instability near 
  periastron.   A mass ejection or wind disturbance may suddenly 
  increase the local density of the primary wind, regions 1 and 4 
  in Fig.\  \ref{ionzones}, thereby changing the colliding-wind 
  parameters and further destabilizing the shocks \citep{kd02,kd05c}.
  In this connection we note a significant discrepancy.  
  The often-quoted mass-loss rate of $\eta$ Car, about 
  $10^{-3}$ $M_\odot$ yr$^{-1}$, would imply a density of the 
  order of $10^{10.5}$ ions cm$^{-3}$ at $r \, = \, 2.5$ AU in a 
  spherical wind.  \citet{pc02}, however, found that
  colliding-wind X-ray models seem to indicate considerably lower
  values.  As \citet{nsmith03} later explained, relatively low densities 
  may exist {\it in low-latitude zones\/} of $\eta$ Car's non-spherical 
  wind if most of the mass usually flows toward high latitudes.   
  Thus, if a mass ejection or wind disturbance occurs during a 
  spectroscopic event, the low-latitude wind may suddenly change 
  from the low-density case exemplified by Pittard \& Corcoran's
  model to a higher-density state.  This affects the 
  \ion{He}{2} $\lambda$4687 emission rate for several reasons that 
  will appear later in this discussion.
  \end{enumerate}
Other scenarios may be possible,  but these three seem most 
obvious.   In the remainder of this section we present a quantitative 
assessment of the \ion{He}{2} $\lambda$4687 problem.  We find that the
observed emission requires either a very large temporary energy 
supply rate, or a highly unusual enhancement of the $\lambda$4687
emission efficiency, or both.

As explained in Appendix \ref{ap_lum}, the STIS data imply a peak
\ion{He}{2} $\lambda$4687 luminosity of about $1.4 \times 10^{36}$
ergs s$^{-1}$.  The uncertainty factor is of the order of 2, but this
is non-Gaussian and an error factor appreciably worse than 3 seems
quite unlikely.  A main conclusion below will be that the \ion{He}{2}
emission was surprisingly bright.  In order to be quite sure of this,
we prefer to lean in the direction of underestimating the
$\lambda$4687 luminosity rather than overestimating it.  Therefore, at
the outset we round the estimate downward to $10^{36}$ ergs s$^{-1}$
or $10^{47.4}$ photons per second.  (Additional allowances will be
made later.)  This is just one emission line, which must be
accompanied by brighter emission in less observable parts of the
spectrum.  The total greatly exceeds $\eta$ Car's 2--10 keV hard X-ray
luminosity.

\citet{steiner} reported a smaller $\lambda$4687 flux at its maximum
(see Section 5 above), and their peak luminosity estimate was only 
about $4 \times 10^{35}$ ergs s$^{-1}$.  This disagreement is not 
crucial in the following analysis.   The most important conclusions
will remain valid even if one adopts a $\lambda$4687 luminosity 
somewhat below their estimate.

\subsection{The Energy Budget for Normal Excitation Processes\label{ebudget}}

Most of the ionic parameters employed here can be found, explicitly
or implicitly, in \citet{osterbrock}.  If \ion{He}{2} $\lambda$4687 
is a recombination line, the peak emission luminosity quoted above
implies about $10^{48.1}$ recombination events per second, only 
mildly dependent on the assumed temperature and density.\footnote{
   Recombination is more efficient than collisional excitation
   for this emission line.  If $T \, < \, 60000$ K in the emitting
   gas, then the collisional excitation rate is hopelessly small by 
   any standard.   At temperatures of the order of $10^5$ K, 
   collisionally excited  $\lambda$4687 emission can be comparable 
   to the recombination emission only if the He$^+$/He$^{++}$ 
   ratio is orders of magnitude larger than the values allowed
   by photoionization, with any photon and electron densities that 
   seem reasonable for this problem.  At $T \, > \, 3 \times 10^5$ K 
   -- e.g., in gas that has recently  passed through a shock --
   collisional ionization of He$^+$ becomes considerably more rapid
   than collisional excitation of $\lambda$4687.   In that case,
   $\lambda$4687 emission carries away an even smaller fraction of the
   total energy than we estimate for the recombination process at
   lower temperatures.}
In order to estimate the maximum plausible efficiency for conversion 
of thermal energy to \ion{He}{2} $\lambda$4687 emission via ``normal''
processes, we make the following assumptions. 
\begin{itemize}  
  \item{Include only helium and hydrogen recombination plus
  bremsstrahlung, and omit excitation of heavy ions and 
  expansion cooling. This simplification will lead to an 
  overestimate, not an underestimate, of 
  the \ion{He}{2} $\lambda$4687 efficiency.}
  \item{The relevant gas is hotter than 50000 K.  In reality,
  lower temperatures may occur if additional processes dominate 
  the cooling;  but in that case the fraction of energy which escapes
  as \ion{He}{2} recombination radiation is reduced.}
  \item{The mass fraction of helium is not much larger 
  than 50\% \citep{davidsonhumphreys1997,hillieretal01}.}
  \item{He$^+$ is ionized mainly by photons between 54.4 and 550 eV;  
  and we assume that all such photons are absorbed by this process 
  rather than escaping.   Most photons above 550 eV either escape, 
  or are converted into 54.4--550 eV emission following absorption 
  by nitrogen and other heavy elements. }
  \item{Temporarily neglect the radiative excitation processes 
  discussed in Section \ref{radexcite} below.}
\end{itemize}

In these circumstances, we calculate that \ion{He}{2} $\lambda$4687 
emission accounts for less than 0.6\% of the escaping radiative energy.   
This result is unsurprising, since $\lambda$4687 accounts for only about 
1\% of the total \ion{He}{2} recombination emission in a conventional 
high-excitation nebula.  We emphasize that 0.6\% is only an upper limit,
and the true efficiency is probably much lower for reasons noted above.  
{\it Even if we adopt this optimistic value it implies an overall 
energy supply of at least $10^{38.2}$ ergs s$^{-1}$, or more than 
$40000 \, L_{\odot}$.\/} If this originates in shocks it is a formidable 
requirement, exceeding the total kinetic energy flow usually quoted 
for $\eta$ Carinae's entire primary wind.  According to \citet{wings04}, 
other observations appear to suggest the same order of magnitude for the 
soft X-ray luminosity during the 2003 spectroscopic event.  Processes
discussed in Section \ref{radexcite} may allow the stellar 
UV radiation field to contribute, thereby reducing the X-rays required,  
but they need special conditions.  Before exploring them, let us review 
the requirements for a model with less than 0.6\% efficiency.

Can the wind of a hot companion star provide most of the relevant 
energy?  Its  speed must be about 
3000 km s$^{-1}$ to account for the observed 2--10 keV 
X-ray spectrum \citep{xrayspec1,pc02,xrayspec2,hamaguchi04}.
A kinetic energy output of $10^{38.2}$ ergs s$^{-1}$ would
thus require a mass-loss rate 
$\sim \; 10^{-4.3}$~$M_\odot$ yr$^{-1}$.  But this is surely an
underestimate;  expansion and escaping  X-rays above 2 keV, rather
than soft X-rays, should account  for most of the post-shock cooling,
and part of the wind escapes to large radii without passing through
the main shock  front.  Therefore, in a simple model of this type the
peak  observed \ion{He}{2} $\lambda$4687 brightness implies an impressive
secondary mass-loss rate of more than  $10^{-4}$ $M_\odot$ yr$^{-1}$.  
Even if we have overestimated the peak $\lambda$4687
brightness by a factor of 3 or 4, the remaining
deduced rate of $\sim 10^{-4.5}$ $M_\odot$ yr$^{-1}$ would 
be extraordinary for a hot massive star.  Evidently this
type of theory requires the extremely unusual primary star to 
have a very unusual companion, without any clear evolutionary 
or physical reason.   Moreover, a secondary star with such 
a fast and massive wind must be very hot and luminous, and 
should therefore produce far more ionizing UV photons than 
the primary does, including photons capable of ionizing 
He$^0$.  Although  we do not yet have enough data for 
a formal calculation, one would expect such an object to 
photoionize the inner parts of the Homunculus nebula, 
causing hydrogen and \ion{He}{1} emission, [\ion{Ne}{3}], 
etc., brighter than the modest amounts observed there.  
We plan to address this detail in a future paper.   

In the above assessment, we consciously trimmed the 
values to {\it favor\/} the hypothesis being tested.  
The adopted peak $\lambda$4687 luminosity of
$10^{36}$ erg s$^{-1}$ is, strictly speaking, about
30\% less than we actually estimated in Appendix \ref{ap_lum};
the assumed conversion efficiency of 0.6\% is most 
likely too high by a factor of 2 or 3;
and we applied a brightness reduction factor  
of 3 or 4 merely for the sake of moderation.  If one repeats the
exercise with the peak flux we observed and the ``most likely''
values, then the required mass-loss rate for the secondary star
becomes roughly $10^{-3.5}$ $M_{\odot}$ yr$^{-1}$ or more instead of
$10^{-4.5}$ $M_{\odot}$ yr$^{-1}$. 

\citet{steiner} favored the fast secondary wind as the main power
source for \ion{He}{2} emission, even with a mass loss rate of only 
$10^{-5}$ $M_{\odot}$ yr$^{-1}$.  They assumed that {\it most\/}
of the secondary wind's kinetic energy is converted to 
He$^{+}$-ionizing photons between 54 eV and 100 eV, and they also
invoked a higher-than-normal efficiency for $\lambda$4687
recombination emission.  Both assumptions require unspecified physical
processes.\footnote{
  The first of these two assumptions was based on an extrapolation
  downward from X-ray energies, using a steep power-law energy
  distribution $f_{\epsilon} \sim {\epsilon}^{-2.7}$.  The exponent
  was estimated from a figure showing Pittard \& Corcoran's
  calculations for the range 1.2--10 keV, i.e., far above 100 eV.
  There is no stated theoretical reason for the spectrum to
  follow such a power law below 1 keV.}
In standard colliding-wind calculations for $\eta$ Car 
(e.g., \citet{pc02}), most of the secondary wind energy is 
accounted for by expansion cooling and moderate-energy X-rays, 
rather than photons below 100 eV.  Applying normal efficiency  
factors, one predicts a \ion{He}{2} $\lambda$4687 luminosity 
considerably below the value Steiner and Damineli quoted -- 
as implied by our analysis presented above.  Their model 
also has a geometrical difficulty.  For reasons noted in 
Section \ref{eclipses} below, those authors concluded that most
\ion{He}{2} emission originated in the acceleration zone of the
secondary wind, region 7 in Fig.\ \ref{ionzones}.  That small locale,
however, cannot intercept more than a few percent of the
He$^{+}$-ionizing  photons produced near the much larger shocked
regions.  In other words, such a model entails a geometrical
efficiency factor smaller than 0.1, not included in any of the
calculations.

The 3000 km s$^{-1}$ secondary wind has other disadvantages for the 
present purpose.  With such a high wind speed, most of the radiation
created near a shock normally has photon energies well above 500 eV, 
very inefficient for ionizing He$^{+}$.  Moreover, the secondary wind 
is expected to be relatively steady because no proposed tidal or 
radiative effects would appreciably alter it, even near periastron. 
This fact precludes some useful effects that can be invoked for the 
more sensitive primary wind (see below).  In summary, based on a 
combination of factors sketched above, the secondary wind is not 
likely to be the main power source for the observed 
\ion{He}{2} $\lambda$4687 emission.   

The primary star's slower, denser wind provides a more likely energy 
supply, soft X-ray supply, and location for the He$^{++}$ region.  
With velocities in the range 300 to 1000 km s$^{-1}$, it produces 
characteristic shock-front temperatures
$kT \lesssim 1$ keV, most likely 200--400 eV, suitable for
ionizing He$^+$.  Steady mass loss at a rate of $10^{-3}$
$M_\odot$ yr$^{-1}$ \citep{cox95,davidson95,hillieretal01} is not
adequate for our purpose, but several effects may improve the
situation.  For instance, there are independent reasons to suspect
that a rapidly growing density enhancement occurs in low-latitude
zones of the primary wind during a spectroscopic event
\citep{zanella84,kd99,kd02,kd05c,nsmith03}.   Our analysis of the STIS
data indicate that roughly $10^{53.5}$ \ion{He}{2} $\lambda$4687
photons were emitted during a 60-day 
interval in 2003, requiring a total input energy of the order of
$10^{44.7}$ ergs if we assume an efficiency of 0.3\%.  This equals the
kinetic energy of $10^{-4}$ $M_\odot$ moving at 700 km s$^{-1}$.  A
temporary ``extra'' mass-flow rate of  several times  
$10^{-3}$ $M_{\odot}$ yr$^{-1}$, passing through 
shock fronts for several weeks, would thus produce the observed 
amount of $\lambda$4687 emission.  Moreover, if the velocities
are unsteady, a complex structure of unstable shocks can form 
around the star, not just at the interface with the secondary wind.
Admittedly the mass-ejection idea is not an entirely satisfying
solution to the energy supply problem for $\lambda$4687, since 
it requires a rather large temporary mass flow rate;  but it 
appears better than any proposed alternative, and the effects 
discussed in Section \ref{radexcite} may help.

Regarding the abrupt peak in the \ion{He}{2} $\lambda$4687 
brightness, note that the apparent time scale may be compressed 
in this scenario.  If velocities tend to increase during 
an event, or if an instability in the shocked gas spreads rapidly,
then the time interval for conversion to \ion{He}{2} emission
can be shorter than the time during which the same energy 
was ejected from the star.  In other words, a moderate-speed 
wind ($V <  1000$ km s$^{-1}$)  has some capacity 
for storing energy which can then be released with suitable 
time and size scales.  Since the hypothetical material moves 
roughly 0.4 AU per day, the likely size scale is of the order
of several AU, comparable to the periastron distance in a binary 
scenario.  The abrupt collapse of $\lambda$4687 emission 
also seems reasonable in a mass-ejection scenario,  because
the ejecta quickly move outward beyond the locale of interest.

Here we have described some characteristics that appear favorable
for a proper quantitative model, which has not yet been developed.
These ideas are broadly consistent with the nature of $\eta$ Car, 
the general appearance of a spectroscopic event, and the requirements 
for explaining the \ion{He}{2} emission.   The energy supply 
appears marginal,  but certain radiative processes may 
enhance the $\lambda$4687 production rate in this type of model.  
They are outlined in the next section.  Later, in Section
\ref{conclusions} we review broader interpretations.


\section{Indirect Amplification by 
    \ion{He}{2} $\lambda$304\label{radexcite}}

In a He$^{++}$ region, trapped $\lambda$304 resonance photons 
(He$^+$ 1s--2p) greatly increase the population of He$^+$ ions 
in the $n = 2$ level.  This fact can enhance \ion{He}{2} 
$\lambda$4687 emission in several ways:
\begin{enumerate}
  \item{Ordinary stellar UV photons with $\epsilon > 13.6$ eV can
    ionize He$^+$ from its $n = 2$ level, thus increasing the
    extent of the He$^{++}$.}
  \item{Trapped hydrogen L$\alpha$ photons may excite He$^+$ ions from
    level 2 to level 4.   (Stellar continuum photons should also
    be considered for the same purpose, see Appendix \ref{apF}.) }
  \item{If the optical depth for the 2--4 transition is appreciably
    larger than unity, then direct decay from level 4 to level 2 becomes
    ineffective because the resulting 1215.2 {\AA} photons cannot easily
    escape.  In that case, unlike a low-density nebula, almost every
    ``successful'' decay from level 4 goes through level 3, producing a
    $\lambda$4687 photon.  This is ``nebular case C'' in the same sense
    as the more familiar ``case B.''} 
\end{enumerate}
Below we investigate these effects.  Note, however, that our idealized
approach tends to overestimate the enhancement factors, for reasons
mentioned later;  a truly realistic analysis would require a far more 
detailed model of the configuration.

Suppose that some region of interest, within a stellar wind, 
is expanding with local velocity gradient $|dv/dx| \, = \, \eta$ 
averaged over directions.  Then, {\it \`{a} la\/} Sobolev, the 
optical thickness in the $\lambda$304 feature along a straight 
path through the region is  
  \begin{equation}\label{rad1}
  \tau_{\lambda304} \ \approx \  
     \frac{ (3.4 \times 10^{-8} \; {\rm cm}^3 \; {\rm s}^{-1}) \; n_1 }
     {\eta} \, ,
  \end{equation}
where $n_1$ is the density of He$^+$ ions in the 1s level at
the location that contributes most strongly to the path integral;
this is fairly well defined if the local turbulent and thermal
velocity dispersion is not too large.  The constant factor in
eqn.\ \ref{rad1} involves the oscillator strength.  In normal circumstances
$\tau_{\lambda304} \, >> \, 1$, and the average escape probability 
for a $\lambda$304 photon is $\Pi \, \approx \, 1/{\tau_{\lambda304}}$ 
per scattering event (see Section 8.5 in \citet{windsbook}). 
{\it Caveat:\/}  By assuming simple homogeneous expansion, we 
are consciously optimizing the entrapment of resonance photons. 
Local instabilities and small-scale velocity gradients may 
decrease the effects discussed below, possibly by large factors.   


Consider the equilibrium densities of He$^+$ ions in levels 1 and 2, 
within the He$^{++}$ region.  Critical parameters are:
  \begin{itemize} 
  \item $R_{c2} \, \approx \, {\alpha}_B \, n_e \, n({\rm He}^{++})$, 
    the effective recombination rate per unit volume.  Here we use
    $\alpha_B$, the standard recombination rate passing through 
    level 2.  Although we do not discuss this detail here, $\alpha_B$
    is a good approximation in the parameter range of interest.
  \item $A_{1c}$, the photoionization rate per ion from 
    level 1 ($\epsilon \, > \, 54.4$ eV).   Note that the photoionization
    cross-section decreases rapidly with photon energy, roughly
    $\sigma(\epsilon) \, \propto \, \epsilon^{-3}$.
  \item $A_{2c}$, the UV photoionization rate from level 2
    ($\epsilon \, > \, 13.6$ eV);  this depends on the Lyman continuum 
    of the primary star, the secondary star, or both.
  \item $A_{21} \, \approx \, 7.5 \times 10^9$ s$^{-1}$,  the 
    $\lambda$304 decay rate.  Here ``level 2'' includes both 2s and 
    2p, since collisional transitions strongly mix them at relevant
    densities. 
  \end{itemize} 
Other effects, e.g.,  collisional de-excitation from level 2 to 
level 1, are negligible.  The equilibrium equations 
then lead to the following densities:  
  \begin{equation}\label{rad2}
  n_1 \ \approx \ ( \frac{\Pi A_{21}}{\Pi A_{21} + A_{2c}} )
          \,  \frac{ R_{c2} }{ A_{1c} }
      \ \approx  \ 
          ( \frac{ A_{21} }{ A_{21} + A_{2c} \, \tau_{\lambda304} } )
          \,  \frac{ R_{c2} }{ A_{1c} }
  \end{equation}
and
  \begin{equation}\label{rad3}
   n_2  \ \approx \ \frac{ A_{1c} }{ \Pi A_{21} } \, n_1  \
     \approx \ \frac{ A_{1c} \, \tau_{\lambda304} }{ A_{21} } \; n_1 \, . 
  \end{equation}
If we define a dimensionless quantity
   \begin{equation}\label{rad4}
   \xi  \  =  \  \frac{ (3.4 \times 10^{-8} \; {\rm cm}^3 \; {\rm s}^{-1}) 
      \, R_{c2} \, A_{2c} }{ \eta \, A_{21} \, A_{1c} }  \;  ,
   \end{equation}
then eqns.\ \ref{rad1}, \ref{rad2}, and \ref{rad3} together imply 
   \begin{equation} \label{rad5}
   n_1 \ \approx \  \frac{2}{(1 + \sqrt{1 + 4 \xi})}  \; 
      \frac{R_{c2}}{A_{1c}} \;  , 
   \end{equation}
   \begin{equation}\label{rad6}
   n_2 \ \approx \ { \frac{4 \, \xi}{(1 + \sqrt{1 + 4 \xi})^2} } \; 
       \frac{R_{c2}}{A_{2c}} \;  .
   \end{equation}
With these densities, the optical depths in the 1--2 and 2--4 
transitions turn out to be
   \begin{equation}\label{rad7}
   \tau_{\lambda304}  \  \approx   \ 
        ( \frac{ 2 \xi }{ 1 + \sqrt{1 + 4 \xi} } ) \;
        \frac{ A_{21} }{ A_{2c} } \; ,
   \end{equation}
   \begin{equation}\label{rad8}
   \tau_{\lambda1215}  \  \approx  \
        \frac{1.13 \, A_{1c}}{A_{21}} \; {\tau_{\lambda304}}^2 \; .
   \end{equation} 
We shall find that $\xi \, \ga \, 1$ is required for strong 
enhancement of $\lambda$4687 emission.  This parameter depends
on location within the He$^{++}$ region;  $\tau_{\lambda304}$
and $\tau_{\lambda1215}$ are ``local'' quantities because they 
refer to the local Doppler velocity in the expanding medium, in the
standard Sobolev manner.

Now let us review the enhancement effects.  If $\xi \, > \, 2$,
then $A_{2c} \, > \, \Pi \, A_{21}$ in eqn.\  \ref{rad2}, and most 
of the photoionization is caused by UV acting on level 2.
In effect the $\lambda$304 photons amplify the extent
of the X-ray ionization, by tapping the stellar UV radiation 
field.  The local amplification factor is 
   \begin{equation} \label{rad9}
   q_1  \   =    \  
      \frac{ A_{1c} n_1  + A_{2c} n_2 }{ A_{1c} n_1 }  \ \approx \ 
      \frac{ 1 + \sqrt{1 + 4\xi} }{ 2 } \; .
   \end{equation} 
Next, one finds that $\tau_{\lambda1215}$, the optical depth
for He$^+$ 2--4 transitions, can be substantial.  Then 
``case C'' rather than ``case B'' describes the He$^+$ 
recombination cascade as mentioned earlier;  so we have
a second enhancement factor for the \ion{He}{2} $\lambda$4687 
production efficiency compared to standard low-density 
nebular formulae.  An adequate approximation is 
   \begin{equation} \label{rad10}
   q_2  \  \approx  \
   \frac{ 1 + 0.44 \, \tau_{\lambda1215} }
       { 1 + 0.26 \, \tau_{\lambda1215} } \; . 
   \end{equation}
Meanwhile, trapped hydrogen L$\alpha$ photons with suitable
Doppler shifts can excite He$^+$ from level 2 to level 4, with 
rate $A_{24}$.  Since $\tau_{\lambda1215}$ is large wherever 
this effect is significant,  we can assume that a 
$\lambda$4687 photon results from every such event.   The 
production rate per unit volume is therefore $n_2 \, A_{24}$.  
A useful way to write this, using eqn.\ \ref{rad6} for $n_2$, is
    \begin{equation}\label{rad11}
    n_2 \, A_{24}  \  =  \  q_3 \, R_{c2} \; ,
    \end{equation}
where
    \begin{equation}\label{rad12}    
    q_3   \   \approx   \     
         \frac{4 \, \xi}{(1 + \sqrt{1 + 4 \xi})^2}  \; 
        \frac{A_{24}}{A_{2c}} \;  .
    \end{equation}
We estimate $A_{24}$ in Appendix \ref{apE}. 


Taking these effects into account, the ratio of \ion{He}{2}
$\lambda$4687 photons to X-ray photoionization events is
approximately
    \begin{equation}\label{rad13}
    Q_{\rm eff}  \  \approx  \  (0.2 \, q_2 + q_3) \, q_1 \; , 
    \end{equation}
where the factor 0.2 is the corresponding production efficiency 
in the standard low-density case where $\xi  <<  1$. At a given 
location in the He$^{++}$ region, $Q_{\rm eff}$ can be estimated 
from eqns.\ \ref{rad9}, \ref{rad10}, and \ref{rad12}. 


Since $Q_{\rm eff}$ is a local quantity, in order to take a valid
average we must explore models of the He$^{++}$ region. Fortunately 
a simplified geometrical structure is adequate for plausibility 
assessments.  Imagine, for example, a configuration illuminated from 
one side by soft X-rays -- crudely representing zone 4 in 
Fig.\ \ref{ionzones}.  Because the lowest-energy photons tend to be 
absorbed first, the ionization rate $A_{1c}$ decreases rapidly with 
increasing depth into the region.  With any reasonable ionizing 
spectrum, $A_{1c}$ has a stronger gradient than the other 
quantities in eqn.\ \ref{rad4};  therefore $\xi$ increases with depth.  
Thus the $\lambda$4687 enhancement factors tend to be largest 
in regions where only the higher-energy photons penetrate, 
typically $\epsilon \, \ga \, 250$ eV.   Provided that the
gas density is not too strongly correlated with depth in the
region, local values of $\xi$ and $Q_{\rm eff}$ are essentially
determined by the average energy $\epsilon$ of photons being 
absorbed there;  consequently, the large-scale shape 
of the He$^{++}$ region -- plane-parallel, convex,
concave, etc.\ -- does not strongly affect the main results.  
Therefore, idealized plane-parallel models appear adequate 
to estimate $Q_{\rm eff}$ within a factor of two or most likely 
better.  We calculated such models for $\eta$ Car with the following 
assumptions:
  \begin{itemize}  
  \item{The region of interest is located near $r \sim 2.5$ AU 
    relative to the primary star.  This does not appear explicitly
    in the formulae, but it influences our choices of the
    other parameters.  In a binary model, $r \sim 2.5$ AU is 
    nearly optimum for the enhancement effects;  smaller values 
    of $r$ would require excessive orbital eccentricities and 
    very rapid periastron passage, while larger $r$ implies 
    smaller densities (see below).}
  \item{The expansion parameter is $\eta \, \approx \, 
    0.8 \, V_{\rm wind} / r \, \approx \, 10^{-6}$ s$^{-1}$.  
    The true value cannot be appreciably smaller than this 
    near $r \, \sim \, 2.5$ AU in the expanding wind, while larger 
    values inhibit the $\lambda$304 entrapment by 
    decreasing the $\xi$-values.\footnote{
      Incidentally,  if the spatial thickness of the He$^{++}$ 
      region is less than about 
      $(5 \; {\rm km} \; {\rm s}^{-1})/\eta \; \sim \; 0.03$ AU, 
      then the resonance photon escape probability $\Pi$ depends
      on the turbulent and thermal velocity dispersion, not on 
      the expansion rate $\eta$.  In that case one must use
      formulae in Section IV of \citet{dn79} rather than
      $\Pi \, \approx \, 1/\tau$.  This is one reason why
      the calculations used here tend to overestimate   
      the enhancement effects.  However, most of the 
      $\lambda$4687 enhancement occurs in ``deep'' zones of
      the He$^{++}$ region, whose characteristic sizes are
      sufficiently large.}
    We shall say more about parameter dependences later below.}
  \item {The electron density is 
    $n_e \, = \, (10^{10.5} \; {\rm cm}^{-3}) \, \mu$,  where 
    $\mu$ is an adjustable parameter.  Note that $\mu$ is 
    of the order of unity in a wind-disturbance model but
    $\mu \, \sim  \, 0.25$ in the \citet{pc02} 
    colliding-wind X-ray model.   We also assume that the helium 
    density is $0.15 \, n_e$,  appropriate for $\eta$ Car.}
  \item{Consistent with the adopted densities, the local 
    He$^{++}$ recombination rate per unit volume is 
    $R_{c2} \, = \, (10^8 \; {\rm cm}^{-3}) \, \mu^2$.
    This is appropriate for a gas temperature around 30000 K;
    and different temperatures can be represented by 
    small adjustments to $\mu$.   Inhomogeneities in the
    wind tend to increase the  volume-averaged value
    of $\mu^2$ but they also increase local values of $\eta$;
    here we ignore this detail because other uncertainties
    are worse.}
  \item{The incident energy flux of photons between 54 eV
    and 550 eV is 
    $\int F_{\epsilon} d\epsilon \, = \, (5 \times 10^{21}
    \; {\rm eV} \; {\rm cm}^{-2} \; {\rm s}^{-1}) \, \chi$, 
    where $\chi$ is another adjustable parameter which we
    take to be unity in most of our calculations.  If 
    $\chi \, = \, 1$, the total flux between  54 eV
    and 550 eV is $10^{37}$ ergs s$^{-1}$ in an area of 5.6 AU$^2$.}
  \item{The assumed spectrum between 54 and 550 eV has shape
    $F_{\epsilon} \, = \, {\rm constant}$,  while photons
    above 550 eV are absorbed by nitrogen instead of
    helium (and, therefore, may be converted to 
    lower-energy photons which are included in our $\chi$
    parameter).  A more realistic spectrum shape would 
    emphasize lower photon energies, thus decreasing the 
    $\lambda$4687 enhancement effects.}
  \item{We assume that the rate for ionization from $n = 2$ 
    by stellar UV photons is 
    $A_{2c} \, \approx \, 10^{4.1}$ s$^{-1}$, which seems 
    realistic for either a binary or a single-star model
    (see, e.g., \citet{hillier05}). }
  \item{$A_{21} \, = \, 7.5 \times 10^9$ s$^{-1}$ for He$^+$, 
    a standard atomic parameter.}
  \item{$A_{24}$, representing excitation by trapped hydrogen
    L$\alpha$ photons, is given by formula \ref{rad4} in Appendix 
    \ref{apE}.  We also calculated models with $A_{24} \, = \, 0$.}
  \end{itemize}

\begin{figure}
\figurenum{9}
\label{newfig}
\includegraphics[scale=.5]{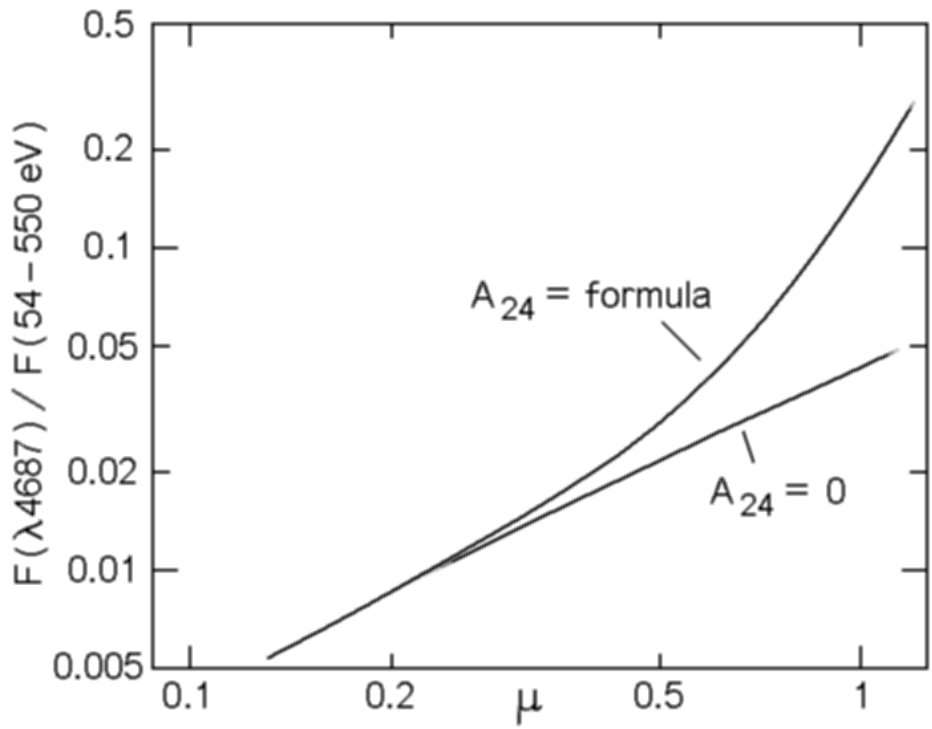}
\caption{The ratio of \ion{He}{2} $\lambda$4687 to
  the input soft X-ray flux as a function of $\mu$.  The upper curve
  includes excitation by Lyman $\alpha$ and the lower curve omits that
  factor.}
\end{figure}

Figure \ref{newfig} shows the resulting ratio of 
\ion{He}{2} $\lambda$4687 luminosity to the input soft X-ray 
luminosity, as a function of $\mu$.   This is a ratio of
energy fluxes, not photon numbers.
The upper curve includes excitation by L$\alpha$ as estimated
in Appendix \ref{apE};  the lower curve omits that process. 
At least in principle, this figure suggests that $\lambda$304 
entrapment can induce an impressive amount 
of $\lambda$4687 emission, possibly more than 10\% of the soft 
X-ray luminosity if $\mu \, \ga \, 1$.  (In that case, of course, 
most of the energy supply comes from L$\alpha$ and the stellar 
UV radiation field, not from the X-rays.  The latter are necessary 
to initiate the process, but then the effects discussed above 
amplify the results by large factors.)   Equally important, 
{\it these processes are not strong enough to explain the 
observed $\lambda$4687 emission if the relevant
density is low,\/} $\mu \, \la \, 0.25$.  At moderately low
densities the emission efficiency 
$F(\lambda4687)/F({\rm soft \; Xrays})$ 
can be of the order of 0.01, which, though much higher than one 
would get without the enhancement effects, is not adequate to 
explain the observations -- see Section \ref{theorysec}.  
Thus the effects discussed above can play 
a major role in a dense wind-disturbance or mass-ejection 
scenario as discussed in Section \ref{theorysec}, but 
they are of little help in the lower-density colliding-wind 
model proposed by \citet{pc02}.\footnote{
   The enhancement effects are weak in the secondary wind's 
   acceleration zone, region 7 in Fig.\ \ref{ionzones}, which 
   \citet{steiner} proposed as the \ion{He}{2} emission region.  
   One reason is that the expansion parameter $\eta$ is large 
   there. }

Can the other variable parameters $\eta$, $\chi$, and $A_{2c}$   
alter this assertion?  Within the plausible ranges, we find
that $F(\lambda4687)/F(54 \; {\rm to} \; 550 \, {\rm eV})$ is 
approximately proportional to $\eta^{-0.5} \, \chi^{-0.45}$.  
The effect of the UV photoionization rate $A_{2c}$ is more 
complex:  For $\mu \, \approx \, 1$ the $\lambda$4687 production 
rate is roughly proportional to 
$A_{2c}^{-0.7}$, for $\mu \, \approx \, 0.5$
the precise value of $A_{2c}$ scarcely affects the results,
and for $\mu \, \la  \, 0.25$ the enhancement factors 
increase with $A_{2c}$;  but the enhancement is insufficient
in that case anyway.   We also calculated some models
with a pseudo-thermal factor $\exp(-\epsilon/500 \, {\rm eV})$
in the input spectrum, and found $\lambda$4687 efficiencies 
about 20\% less than in the models outlined
above.  In summary, $\mu$ is the dominant parameter.

Realistic $\lambda$4687 enhancement factors may be far less than 
we estimated above, because instabilities in the wind can produce 
local velocity gradients that help resonance photons escape.  
Moreover, any unrecognized process that destroys $\lambda$304 
or L$\alpha$ photons would also reduce the effects discussed here.  
Thus, in a sense we have derived {\it upper limits\/}.

The essential result for $\eta$ Car is that $\lambda$4687 emission can 
be substantially enhanced only in a region where the electron and ion 
densities exceed, roughly, $10^{10.2}$ cm$^{-3}$ ($\mu \gtrsim 0.5$).  
This may reduce the energy supply difficulty for the type of model
we advocated above, with temporary high densities.  The observed rapid 
growth and decline of the feature around its maximum seem to make sense 
in this case, because the amplification factor is quite sensitive to 
the gas density.  However, the same effects do not provide sufficient 
amplification at densities below $10^{10}$ cm$^{-3}$ ($\mu < 0.3$).  
Therefore they seem unlikely to account for $\lambda$4687 in the
lower-density type of colliding wind model calculated by \citet{pc02}.

Conceivably the \ion{He}{2} $\lambda$4687 might be excited
by stellar photons near 1215 \mbox{\AA}, absorbed by He$^+$ ions
in the $n = 2$ level. According to a brief assessment
sketched in Appendix \ref{apF}, this phenomenon may account for
an equivalent width of the order of 0.1 \mbox{\AA} but not
much more -- so it, too, appears to be inadequate in
a low-density model.

Unfortunately these quantitative results depend too strongly
on the geometrical assumptions for us to feel entirely
confident about them.   For reasons noted above, most likely 
we have {\it overestimated\/} the enhancement factor. (This 
is especially true for region 6 in Fig.\ \ref{ionzones}.)  
On the other hand, since the basic parameters are poorly 
established, an inventive theorist may be able to construct 
a low-density model by carefully tailoring the assumptions.  
Obviously this problem needs more work, including realistic 
simulations with the appropriate radiative processes and 
gas-dynamical instabilities.  


\section{Implications\label{conclusions}}

\subsection{Concerning Eclipses\label{eclipses}}

Even without emission rate analyses, our measurements of \ion{He}{2} 
$\lambda$4687 are difficult to reconcile with the 
eclipse explanation for $\eta$ Car's spectroscopic events advocated
by \citet{ecl98}, \citet{pc02}, \citet{steiner}, and others.  
Here the term ``eclipse'' means that an event occurs when
the secondary star and X-ray region move behind the far side
of the primary stellar wind.  Some of the arguments listed below are
individually not strong, but collectively they indicate that the
emission did not behave as one would expect in a straightforward
eclipse model. 

\begin{enumerate}
\item{For reasons discussed earlier, the \ion{He}{2} emission and 
the X-rays probably originated in the same locale;  no one has 
suggested a plausible alternative.  Yet the $\lambda$4687 feature 
peaked much later than the 2--10 keV X-rays -- in fact, at a time 
when the observable X-ray flux had already fallen to a small fraction 
of its maximum.  As M.\ Corcoran has remarked (priv.\ comm.),
in an eclipse model this implies that the X-rays at that time were 
strongly absorbed by the intervening primary wind, but the \ion{He}{2} 
$\lambda$4687 emission was not.   The wind, however,  is not fully 
transparent near 4680 {\AA}; column densities sufficient to thoroughly 
block X-rays above 6 keV would also cause visual-wavelength Thomson 
scattering with optical depths of the order of unity \citep{xabs}.  
More important for our argument, the apparent \ion{He}{2} emission 
{\it rose dramatically\/} while the observed 2--10 keV X-rays fell.  
This fact has not been explained in an eclipse model,  but is 
relatively straightforward in non-eclipse models (see Section
\ref{sec72}).} 
\item{The \ion{He}{2} $\lambda$4687 feature disappeared quite
rapidly after its peak (Section \ref{gonequick} above, and
\citet{steiner}).  During that time interval the secondary star
moved only about 1 AU along its relative orbit -- comparable to 
the size of the likely \ion{He}{2} emission region (Fig.\
\ref{ionzones}).   The intervening ``object'' in almost every 
proposed eclipse model is the polar wind of the primary star, 
not the star itself.   Therefore, if the rapid disappearance of 
the \ion{He}{2} feature was an eclipse phenomenon, the primary 
wind must be unexpectedly sharp-edged in some sense.  
This statement is too ill-defined to be a strong argument against
the eclipse idea, but it does pose an additional constraint
-- while, by contrast, a disappearance timescale of 5--10 days
is unsurprising if it was due to other causes as we suggest
below. }
\item{As mentioned in Section \ref{gonequick} above, VLT/UVES
observations of location FOS4 in the Homunculus showed a 
disappearance of the 4680 {\AA} emission at about the same
time as in our direct view -- a little earlier, perhaps, but 
essentially at the same time \citep{stahl05}.  Since those 
observations represent a reflected pole-on view of 
$\eta$ Car, there is no reason for them to show the postulated 
eclipse.  Thus there are two choices:  (a) If the rapid 
disappearance of \ion{He}{2} emission in our direct view was 
due to an eclipse, then a separate explanation must be found 
for the VLT/UVES results, with a fortuitous near-coincidence 
in timing. (2) Alternatively, the \ion{He}{2} $\lambda$4687 
behavior  was dominated by other effects which applied in both 
directions, and had little to do with the X-ray eclipse.  
Either choice is unpalatable if one desires a simple model.  (In this
connection, note that most proposed eclipse models require the event
to occur very near periastron, in which case some effects depend on
the eclipse but others depend on periastron passage.  This coincidence
of eclipse and periastron requires an entirely fortuitous special
orientation of the orbit.) }
\item{At its maximum, the $\lambda$4687 emission peak had a Doppler 
velocity near $-450$ km s$^{-1}$ (Fig.\ \ref{evol4686b}).  In an
eclipse model, however, the emission region at that time should 
have been on the far side of the primary star,  where the 
apparent Doppler velocity of the primary wind is {\it positive.\/}   
In order to reconcile this with the observed \ion{He}{2} profile 
\citet{steiner} concluded that the \ion{He}{2} emission comes from
region 7 in the secondary wind rather than region 4 in the primary wind
(Fig.\ \ref{ionzones}).  The resulting geometric efficiency factor ($<
0.1$, noted in Section \ref{ebudget}) makes that
suggestion difficult if not impossible on 
energy-supply grounds.  Region 4 has far better parameters for \ion{He}{2} 
emission.  In a non-eclipse model the observed velocity seems more
or less reasonable for Region 4, because there is no need for it to be
located on the far side of the primary star at the critical time --
see Section \ref{sec72} below.  }
\item{According to our analysis, the observed maximum 
\ion{He}{2} $\lambda$4687 luminosity appears to indicate a 
higher primary-wind density than \citet{pc02} found suitable 
for their eclipse model.  The higher density is needed
to provide an adequate energy supply for soft X-rays 
(Sec.\ \ref{theorysec}) and to allow radiative enhancement 
processes (Sec.\ \ref{radexcite}).}
\end{enumerate}
Evidently, then, the eclipse hypothesis requires one or more
additional phenomena as well as a special orbit orientation.  

\subsection{Models Without Eclipses\label{sec72}}

There is an alternative scenario that does not depend on an 
eclipse.  Others are conceivable, some of them with only a 
single star, but this one is easiest to describe.
Assume that a hot companion star follows a highly eccentric orbit
whose approach-to-periastron and periastron itself are not on the 
far side of the primary.  Colliding winds produce the observed 
2--10 keV X-rays in the usual manner \citep{xrays92,pc02}. 
As the secondary star approaches periastron, the shocked gas 
becomes denser;  this increases the radiative fraction 
of cooling there, with two well-known results seen in the
observations \citep{xrays2,xraycurve}:
The X-ray luminosity increases and the shock surfaces 
become somewhat unstable.  Next, suppose that near periastron  
{\it the secondary star's tidal and radiative forces induce a 
major disturbance in the primary star's inner wind\/}
\citep{kd99,kd02,kd05c}.  This may be the mass 
ejection proposed by \citet{zanella84}, or a temporary 
alteration of latitude dependences suggested by 
\citet{nsmith03}, or some combination of both; and it need not 
have either spherical or axial symmetry.  Anyway the relevant 
gas density increases rapidly, perhaps by a large factor.  

As \citet{kd02} proposed, one result may be a catastrophic breakup
of the large-scale shock structure.  High densities imply 
high radiative efficiencies which promote the instabilities 
described by \citet{xrays92}.  \citet{pc02} remarked that in 
their calculated models for $\eta$ Car, the primary-wind side 
of the shock structure is unstable but the secondary side, 
where observable X-rays above 1 keV originate, is stable by 
moderate factors.  A serious density increase would reduce 
the stability margin of the latter and may even 
destabilize it.\footnote{
   Near periastron, their parameter $\chi$ is of the order of 10 on 
   the secondary side;  instability tends to arise if $\chi \lesssim 1$.  
   Based only on their models, one might argue that $\chi$ would still 
   exceed unity after, say, a factor-of-ten density increase.  However, 
   those models are simplified in various respects, their quantitative 
   assumptions are debatable, and the stability criterion is rather 
   ill-defined;  so our suggestion is not fundamentally implausible.  
   See also an ``empirical'' remark a little later in the text. }
The entire structure would thus become more chaotic and more complex 
than it was earlier.  The resulting combination of oblique 
angles of shock incidence and multiple subshocks tends to reduce 
the average temperature in the shocked gas.  The 2--10 keV X-rays 
then fade and disappear for two reasons:  (1) The overall production 
spectrum softens.  (2) Column densities through the freshly ejected 
gas are of the order of $10^{24}$ ions per cm$^2$, comparable 
to those invoked in an eclipse model, even though the X-ray 
source region is not on the far side of the primary. 

The observed X-rays became remarkably unsteady before the 1997 and 
2003 events \citep{xrays2,xraycurve}.  Thus it seems fair 
to deduce, empirically and without reference to detailed models,
that the pre-event shock structure was susceptible to disruption 
of the type envisioned here.  Given the large-amplitude 2--10 keV 
flaring seen near maximum, it would be surprising if a large 
rapid density increase does {\it not\/} alter the structure. 

In this scenario most of the observed ultraviolet-to-infrared 
spectroscopic changes occur because the additional outflow
temporarily quenches the ultraviolet flux,  as \citet{zanella84}
originally proposed.  This is true even if the hot secondary star 
produces most of the ionizing UV, and the observed delay between 
the 2--10 keV X-ray maximum and the main UV-to-IR event 
(Fig.\ \ref{comptiming}) seems reasonable (see below). We 
have no quantitative model for the UV-to-IR effects, but 
none has been developed for the eclipse hypothesis either.  

A mass ejection or wind disturbance model appears well adapted 
to explaining the \ion{He}{2} $\lambda$4687 emission, for three 
reasons:   (1) Chaotic, vigorously unstable shocked gas can produce 
copious soft X-rays and extreme ultraviolet emission suitable 
for ionizing He$^+$. (2) The outflowing material provides a 
temporary energy supply as noted in Section \ref{theorysec} above.  
(3) Enhanced densities in the primary wind are favorable for the 
radiative amplification processes described in Section \ref{radexcite}. 
This type of model avoids the difficulties that apply to eclipse 
models, as listed in the preceding sub-section.  The delay between 
the observed 2--10 keV X-ray maximum and the \ion{He}{2} maximum 
is particularly noteworthy.  During that interval, we suspect, 
{\it soft\/} X-ray production ($\epsilon < 500$ eV) increased 
as the shock structure became more unstable.  Indeed, in a model of
this type the 2--10 keV X-rays play almost no role in the main event. 
The duration of the wind disturbance must be several weeks, roughly
the time from the 2--10 keV X-ray maximum to the \ion{He}{2}
$\lambda$4687 peak;  this may represent the time interval when the
secondary star is close to periastron in an appropriate dynamical
sense.  The rapid timescale of the $\lambda$4687 decay also seems
reasonable for the last ``extra'' ejecta to exit the vicinity.
A characteristic flow speed 
$\sim$ 500 km s$^{-1}$ and a characteristic size scale $\sim$ 1 AU 
would imply a timescale $\sim$ 3.5 days, comparable to but 
comfortably shorter than the observed times of growth and decline.   

As mentioned above, this type of model requires column densities of
the order of $10^{24}$ cm$^{-2}$, or possibly more, at the time of the
event.  Therefore the optical depth for Thomson scattering of the
$\lambda$4687 emission may be appreciable. Since the emission zone is
not on the far side of the primary star, this does not increase the
energy problem;  much of the scattered light will emerge in our
direction.  However, the line wings near maximum may be strongly
affected by Thomson scattering.

\citet{bish01} has noted that $\eta$ Car's 2--10 keV X-ray
variations are easiest to explain, through most of their 5.5-year
cycle, if the major axis of the eccentric orbit is roughly 
perpendicular to our line of sight;  see Fig.\ 1 in \citet{kd02}.  
A few spectroscopic details observed with HST/STIS support the same
idea \citep{halpha}.  Such an orbit orientation is very different from
the eclipse models cited above but 
it would be consistent with a wind-disturbance model.  If this
orientation is more or less correct, then we expect the secondary star
to move behind the primary wind a few weeks {\em after} the observed event
-- i.e., at a time when it has little obvious effect on the main
observables.  The post-event spectroscopic and X-ray recovery may
depend on emergence from the true eclipse, but detailed  models are
needed to assess this question. 

\subsection{Summary}

The HST/STIS observations confirm that an emission feature
arose near 4680 {\AA} just before $\eta$ Car's mid-2003 
spectroscopic event, and that it originated fairly close to
the central star.  The most likely identification is
\ion{He}{2} $\lambda$4687.  Several details   
influence the theoretical picture.

\begin{enumerate}
\item{Our measurement of the feature's maximum flux is more than twice
  what was reported by \cite{steiner}.}
\item{The relationship between the observed 2--10 keV X-rays
and the \ion{He}{2} emission is complex and indirect (Section
\ref{compothers}).  This is a good example of a seldom-noted circumstance:
For a physical model of $\eta$ Car's spectroscopic events, the UV-to-IR
observables (e.g.\ those shown in Fig.\ \ref{comptiming}) are far more
crucial than the 2--10 keV X-rays.  They represent larger energy
flows, and they represent processes at various locations in the dense
primary wind, rather than merely the shock surface of the secondary
wind.  The 2--10 keV X-ray peak does not coincide in time with the
spectroscopic event.} 
\item{We detected no \ion{He}{2} emission before 2003 or after 2003.5.}
\item{Our analysis indicates that rather high densities, higher than
  in some proposed X-ray models, are needed to provide an adequate
  energy supply and allow radiative enhancement processes.}  
\item{The energy supply and observed line profile strongly imply that
  the emission originates in region 4 (Figure 6).} 
\item{The Doppler velocity of the line peak, roughly $-450$
km s$^{-1}$ at the time of maximum, is difficult to explain in an
eclipse model.}
\end{enumerate}

A wind-disturbance model for $\eta$ Car's spectroscopic events is
logically simpler than an eclipse scenario.  At first sight this
assertion may seem novel, since eclipses are fairly common in
astronomy while the postulated wind disturbance requires a stellar
instability.  However, a number of observations indicate that the
eclipse idea does not suffice by itself;  it needs additional
phenomena to work.  In section \ref{eclipses} we listed some of these
indications related to \ion{He}{2} $\lambda$4687. There are others \citep{kd99,
halpha,kd05c}, and \citet{xraycurve} has recently acknowledged that
some aspects of the 2--10 keV X-ray observations may imply a wind
disturbance.  Thus it now appears that a wind disturbance is required,
{\em even if there is an eclipse}.  Moreover, as noted in subsection
\ref{sec72} and by \citet{kd99} and \citet{kd02},  {\it a substantial
  mass-ejection event would cause the 2--10 keV X-rays to disappear
  with no need for an eclipse\/}.  

A mass-ejection or wind disturbance is potentially significant for 
stellar astrophysics and gas dynamics in general;  it requires some 
undiagnosed surface instability which must depend on the the primary 
star's structure.  For an ordinary object one hesitates to invoke an 
``undiagnosed instability,'' but during the past 200 years $\eta$ Car 
has repeatedly exhibited other phenomena fitting that description.

In view of all the unexplained facts, the emission near 4680 {\AA} 
constitutes a significant and interesting theoretical problem. Like 
many other outstanding puzzles involving $\eta$ Car, it may 
relate to several branches of astrophysics and deserves careful 
attention. 

\section{Acknowledgments}
We are grateful to M. Feast for inspecting Thackeray's spectra of
Eta Carinae for us.  
We thank O. Stahl, K. Weis, and the Eta Carinae UVES Team for
their continued collaboration and sharing of data.
We also thank K. Nielsen and G. Vieira-Kober for providing us with the
reduced and extracted STIS/MAMA spectra.
We acknowledge M. Corcoran for providing many constructive
comments and discussion regarding an early draft. 
Additionally, we are grateful to T.R. Gull and Beth Perriello for
preparing the HST observing plans.  
We also especially thank
J.T. Olds, Matt Gray, and Michael Koppelman at the University of
Minnesota for helping with non-routine steps in the data preparation
and analysis. 

The HST Treasury Project for Eta Carinae is supported by NASA funding
from STScI (programs GO-9420 and GO-9973), and in this paper we have
employed HST data from several earlier Guest Observer and Guaranteed
Time Observer programs.

This work also made use of the NIST Atomic Spectra 
Database\footnote{http://physics.nist.gov/cgi-bin/AtData/main$\_$asd}
and the Kentucky Atomic Line List
v2.04\footnote{http://www.pa.uky.edu/$\sim$peter/atomic/index.html}.

\appendix



\section{Intrinsic Luminosity of the Observed Emission Feature\label{ap_lum}}

Eta Car's large and uncertain circumstellar extinction makes it
difficult to convert the observed \ion{He}{2} $\lambda$4687 flux 
directly to an  intrinsic luminosity.  Fortunately, however, 
the star's intrinsic brightness is known to a useful accuracy, 
independent of the circumstellar extinction.  Therefore the best 
approach is to use the underlying stellar continuum as a calibration 
reference.  Based on theoretical considerations as well as the 
spatial resolution of STIS, we assume that the $\lambda$4687 
emission originates in the dust-free region within about 150 AU 
of the star.  One could estimated the continuum brightness
from a theoretical model of $\eta$ Car's wind, but that approach
seems unsafe to us because it requires a number of implicit, 
mostly unexpressed assumptions.  We prefer to use $\eta$ Car's 
brightness observed 200 to 400 years ago, before the Great Eruption 
produced the circumstellar extinction.  This approach requires 
two critical assumptions: that the star's luminosity has not 
changed much, and that circumstellar extinction was small before 
1800.  Both seem consistent with all available theoretical 
and observational data.

Before 1800 the star varied between fourth and second magnitude
at visual wavelengths, with (probably) little change in total
luminosity.  In the fainter state, the continuum resembled a
hot star with a relatively transparent wind;  and the 
second-magnitude state occurred whenever the wind became 
dense enough to form a photosphere cooler than 9000 K   
\citep{davidsonhumphreys1997}.  
The present-day state of the wind is closer to the hotter case,
but, most likely, with a somewhat increased mass-loss rate;  
therefore the brightness would now be third or fourth magnitude 
if the circumstellar extinction were not present.  We adopt
$V \, \approx \, 3.5$, which implies  $V_0 \, \approx \, 2.0$ when 
corrected for interstellar extinction.  The r.m.s.\ uncertainty 
is perhaps 0.5 magnitude, but this is non-Gaussian in the sense 
that an error worse than about 0.8 magnitude seems very unlikely.  
If the intrinsic color is $(B - V)_0 \, \approx \, -0.1$ based 
on the wind's spectroscopic character (this detail does not 
strongly affect our result), and if $D \, \approx \, 2.3$ kpc, 
then $\eta$ Car's intrinsic luminosity per unit wavelength near 
4687\mbox{\AA} is about  
$6 \times 10^{35}$ erg s$^{-1}$ \mbox{\AA}$^{-1}$.  
This is about 40\% larger than Steiner and Damineli's 
calibration, a reasonable agreement in view of the
uncertainties.

Therefore an emission equivalent width of 1 \mbox{\AA} corresponds 
to  intrinsic luminosity 
$L({\lambda}4687) \, \approx \, 6 \times 10^{35}$ erg s$^{-1}$
or $1.4 \, \times \, 10^{47}$ photons per second.  
If this is an {\it over-\/}estimate worse than a factor of 
about two, then some underlying assumption that we share
with \citet{steiner} must be fundamentally incorrect 
-- which would be interesting in itself.

As discussed in Section \ref{sec3}, the STIS data indicate that the 
\ion{He}{2} $\lambda$4687 equivalent width briefly reached 2.4 {\AA} 
around MJD 52813.8 (Table \ref{ewfluxtab}, Fig.\ref{evol4686b}), with
a very broad velocity dispersion.  This value corresponds to
$L({\lambda}4687) \, > \, 1.4 \times 10^{36}$ erg s$^{-1}$, an
impressive amount even if we were to reduce it by a factor of two or
three merely for the sake of moderation. 



\section{L$\alpha$ and the $A_{24}$ Rate\label{apE}}

Hydrogen L$\alpha$ $\lambda$1215.7 with a suitable wavelength
shift can excite He$^+$ from its $n = 2$ level to $n = 4$ 
(1215.2 {\AA}), and subsequent decay to $n = 3$ produces a 
$\lambda$4687 photon.  If we measure wavelength or frequency
in terms of the Doppler parameter $v \, = \, 
(\lambda - \lambda_{{\rm L}{\alpha}})/{\lambda_{{\rm L}{\alpha}}}$,
the He$^+$ 2---4 transition occurs at 
$v_{\, \rm HeII} \, \approx \, -120$ km s$^{-1}$.  Let us represent
the L$\alpha$ radiation field by an equivalent temperature 
$T_{\rm rad}(v)$, such that the photon energy density or average 
specific intensity at $v$ is given by the Planck formula at that 
temperature.  This equivalent temperature is determined by the 
hydrogen population ratio $n_2/n_1$ and by the optical depth
at $v$.  The resulting excitation rate for a He$^+$ ion in
its $n = 2$ level is   
   \begin{equation}\label{F1}
   A_{24}({\rm He}^+)  \  \approx  \  (5 \times 10^8 \; {\rm s}^{-1}) \; 
    \exp( -\frac{118400 \; {\rm K}}{T_{\rm rad}(v_{\, \rm HeII})} ) \, ,  
   \end{equation} 
based on the transition probability $A_{42}$ for He$^+$, 
statistical weights, and transition energy.  Here we estimate
$T_{\rm rad}(v_{\, \rm HeII})$ in the inner wind of $\eta$ Car.  


Again considering a region near $r \, \sim \, 2.5$ AU, we use 
the same expansion rate $\eta$ and gas density parameter $\mu$
as in Section \ref{radexcite}, and the following rates for hydrogen:
  \begin{itemize}
  \item $R_{{\rm H},c2} \; \approx  \; {\alpha}_B n_e n({\rm H}^+) \; 
      \approx \; 10^8 \, \mu^2$ cm$^{-3}$ s$^{-1}$,
      the recombination rate through level 2.  This is accidentally
      equal to the value that we used for helium in Section
      \ref{radexcite}, but the details differ.  The temperature
      dependence is modest compared to other uncertainties, and can be
      incorporated in the density parameter $\mu$. 
   \item $A_{{\rm H},1c} \, \sim \, 10^{4.4}$ s$^{-1}$, the  
      photoionization rate from level 1.  This is approximately
      equal to $1.8 A_{2c}({\rm He}^+)$ (Section \ref{radexcite}) because it 
      involves the same UV flux between 13.6 and 24 eV.
   \item $A_{{\rm H},2c} \, \approx \, 10^{5.8}$ s$^{-1}$, the 
      photoionization rate from level 2.  Since this involves 
      mainly the primary star's flux in the Balmer continuum,
      it is less uncertain than $R_{{\rm H},c2}$  and 
      $A_{{\rm H},1c} \,$.  
   \item $A_{{\rm H},21} \, \approx \, 4.7 \times 10^8$ s$^{-1}$,
      the decay rate that produces L$\alpha$.
   \end{itemize}
Collisional transitions mix levels 2s and 2p, other collisional 
transitions are negligible, etc.   Most of the formulae in 
Appendix \ref{ap_lum} apply to the trapped L$\alpha$ photons, except that 
for hydrogen we must replace eqn.\ \ref{rad1} with
  \begin{equation}\label{F2}
  \tau_{\, {\rm L}\alpha} \ \approx \ 
     \frac{ (1.34 \times 10^{-7} \; {\rm cm}^3 \; {\rm s}^{-1}) 
      \; n_{{\rm H},1} }
     {\eta} \, ,
  \end{equation}
and $1.34 \times 10^{-7}$ cm$^3$ s$^{-1}$ must be substituted 
for $3.4 \times 10^{-8}$ cm$^3$ s$^{-1}$ in the formula analogous
to eqn.\ \ref{rad4}. The latter then indicates that    
$\xi_{\, {\rm H}} \, \approx \, 0.7 \, \mu^2$.  For example, in a 
dense model with $\mu \, \approx \, 1$ we find 
$\tau_{\, {\rm L}\alpha} \, \approx \, 360$   and 
$n_{{\rm H},2}/n_{{\rm H},1} \, \approx \, 0.02$ 
which corresponds to an excitation
temperature slightly above 22000 K.  This means that $T_{\rm rad}$
is about 22000 K at any wavelength where L$\alpha$ is optically
thick.  


However, at $v_{\, \rm HeII} \, \approx \, -120$ km s$^{-1}$
the gas is probably {\it not\/} optically thick, since nearly 
all surrounding material has positive $v$ relative to the location
of interest.  The wavelength difference is so large that the 
natural line wing dominates.  Using the appropriate Lorentzian 
profile in the path integral that led to eqn.\ \ref{F2} for the line 
center, one finds an optical depth
   \begin{equation}\label{F3}
   \tau(v_{\, \rm HeII})  \  \approx  \
      - \, ( \frac{\lambda_{{\rm L}\alpha} A_{{\rm H},21}}
          {4 \pi^2 v_{\, \rm HeII}} ) 
      \, \tau_{\, {\rm L}\alpha} 
      \  \approx  \  
      10^{-4.8} \, \tau_{\, {\rm L}\alpha} 
      \, \approx \, 0.006 \, .  
   \end{equation}
Including this factor, $T_{\rm rad}(v_{\, \rm HeII})$ is reduced to  
only about 11400 K.  A practical approximation for the resulting 
He$^+$ excitation rate is 
   \begin{equation}\label{F4}
   A_{24}({\rm He}^+, {\rm L}\alpha) \   \approx  \ 
      (23000 \ {\rm s}^{-1}) \; 
      ( \frac{A_{{\rm H},1c}}{10^4 \ {\rm s}^{-1}} )^{-0.75} \; 
      \mu^3   
      \\   \   \approx  \
      (15000 \ {\rm s}^{-1}) \; 
      ( \frac{A_{2c}({\rm He}^+)}{10^4 \ {\rm s}^{-1}} )^{-0.75} \; 
      \mu^3  \; ,  
   \end{equation}
which agrees with our formulae to an accuracy better than 15\% in the
range $0.4 \, < \, \mu \, < \, 3$,  
$10^{3.9} \ {\rm s}^{-1} \, <  \, A_{{\rm H},1c} \, < \,
  10^{4.9} \ {\rm s}^{-1}$.  Larger values of $\mu$ are 
unlikely in this context, and for smaller values the 
excitation process is relatively weak.   


\section{Excitation by Stellar Radiation near 1215 \mbox{\AA}\label{apF}}

Stellar radiation near 1215.2 \mbox{\AA} can excite He$^+$ from
level 2 to level 4.  The details are very different from
excitation by trapped L$\alpha$ photons, because stellar
radiation arrives from outside the He$^{++}$ region.
As a simple example, imagine a small sub-region with volume
$\Delta V$, illuminated by a beam of continuum radiation 
with flux $F_{\lambda}$.  The gas in $\Delta V$ is expanding
with rate $\eta$ defined in Section /ref{radexcite}.  If the Sobolev-style 
local optical depth $\tau_{\lambda1215}$ is fairly large, 
one can show that the total amount of radiation scattered 
by He$^+$ in volume $\Delta V$ is approximately
$(\lambda \, F_{\lambda})(\eta \, \Delta V / c)$.
This is extracted from a wavelength interval $\Delta \lambda$ 
that depends on the velocity dispersion within $\Delta V$. 
Therefore, if $\tau_{\lambda1215} \, \ga \, 1$ in our 
$\eta$ Car problem, then {\it the total rate of scattered 
photons is proportional to the volume of the He$^{++}$ region.} 
In contrast to the effects described in Section \ref{radexcite}, this
favors low-density cases.  (If the density becomes too low,
however, then $\tau_{\lambda1215}$ becomes small.)

Consider incident radiation from the primary star, or rather
from the opaque inner wind.  Initially let us pretend that its 
luminosity spectrum is a smooth continuum $L_{\lambda}$. 
Suppose that every photon within Doppler interval
$\Delta \lambda \, = \, \lambda \, \Delta v / c$ centered
near 1215 \mbox{\AA}, and within solid angle $\Omega$, is
intercepted by a He$^+$ ion in its $n = 2$ level and
thereby converted to a photon pair at 4687 and 1640 {\AA} 
(the 4--3 and 3--2 decays).  Then, relative to the star's
continuum, one can show that the equivalent width of the 
resulting pseudo-emission line at $\lambda4687$ is
   \begin{equation}\label{G1}
   {\rm EW}_{\lambda}(\lambda4687)  \  \approx  \ 
      (1215 \, {\rm \mbox{\AA}}) \; ( \frac{\Delta v}{c} ) 
      \; ( \frac{\Omega}{4\pi}) \; S  
      \  \approx  \
      (0.4 \, {\rm \mbox{\AA}}) \; 
	  ( \frac{\Delta v}{100 \, {\rm km/s}} ) 
      \; ( \frac{\Omega}{4\pi}) \; S \; ,
   \end{equation}
where
   \begin{equation}\label{G2}
	S  \  =  \  
       \frac{(\lambda \, L_{\lambda})_{\lambda1215}}
            {(\lambda \, L_{\lambda})_{\lambda4687}} \; .
   \end{equation}
For instance, the Planck formula would give $S$ = 0.56 
and 2.2 for $T$ = 15000 K and 20000 K respectively.
Considering that a L$\alpha$ absorption line is formed 
in innermost layers of the wind, it would be difficult 
for $S$ to appreciably exceed 1 in the case of $\eta$ 
Car.  In a colliding-wind binary model, the spherical
coverage factor ${\Omega}/4{\pi}$ is less than 0.1
and the applicable $\Delta v$ is not likely to exceed 
200 km s$^{-1}$ even in a low-density model.  Therefore
the \ion{He}{2} $\lambda$4687 ``emission'' due to this
process may have an equivalent width of the order of 
0.1 \mbox{\AA} near periastron, but not appreciably more.

The hypothetical companion star must be quite hot as
discussed elsewhere;  and, based on the lack of an
impressive photoionized \ion{H}{2} region in the
inner Homunculus, its luminosity cannot be much
more than 10\% that of the primary.  Combining these
factors, its value of $L_{\lambda}(\lambda1215)$
is less than 10\% of the value for the continuum
of the primary (or rather the primary's inner wind),
and most likely less than 5\%.  Therefore, even
though its L$\alpha$ absorptions line is presumably weaker and its
applicable covering factor ${\Omega}/4{\pi}$ might be as large as 0.5
rather than less than 0.1, the secondary star cannot substantially
increase the above estimate.  

L$\alpha$ photons from the secondary wind are
less important in this connection than the
stellar radiation, because the hypothetical
secondary wind is not particularly dense.

Since we have employed several uncertain parameters
here, a model with different results seems possible
though unlikely.  More detailed calculations are
obviously needed.


\clearpage   

\begin{deluxetable}{lcccl}
\tablewidth{0pt}
\tabletypesize{\tiny}
\tablecolumns{5}
\tablecaption{Predicted \ion{He}{2} Line Fluxes\label{heiitab}}
\tablehead{
 \colhead{Wavelength}&\colhead{Relative}&\colhead{Predicted}
 &\colhead{Detection}\\
\colhead{(\mbox{\AA} in
  Vacuum)}&\colhead{Strength\tablenotemark{a}}&\colhead{Flux\tablenotemark{e}}
 &\colhead{Limit\tablenotemark{b}}&\colhead{Notes}}
\startdata 
1640&8.150&601.&5.70&High extinction\tablenotemark{d}\\
2386&0.104&10.3&6.52&High extinction\tablenotemark{c}\\
2512&0.193&14.2&5.81&High extinction\tablenotemark{c}\\
2734&0.257&19.0&5.87&High extinction\tablenotemark{c}\\
3204&0.469&34.6&6.02&Confusion with other features\tablenotemark{c}\\
3797&0.003&0.221&9.26&Blend with Balmer line\\
3925&0.008&0.590&6.96\\
4101&0.015&1.11&8.75&Blend with Balmer line\\
4339&0.026&1.92&6.90&Blend with Balmer line\\
4543&0.036&2.66&7.18\\
4687&1.000&73.8&7.69\\
4861&0.053&3.91&7.82&Blend with Balmer H$\beta$\\
5413&0.080&5.90&16.0\\
6562&0.134&9.88&15.7&Blend with Balmer H$\alpha$\\
10126&0.237&17.5&19.8&High CCD noise\\
\enddata
\tablenotetext{a}{Relative to F(He II 4687\mbox{\AA})=1.0.
  Taken from \citet{osterbrock} for T=40,000 K and
  $4{\pi}j_{\lambda4686}/N_{He^{++}}N_e=3.48 \times 10^{-25}$(erg cm$^{-2}$
  s$^{-1}$).}  
\tablenotetext{b}{Minimum detectable line flux
  ($\Phi_{min}=\left(\frac{f_\lambda}{(S/N)_1}\right){\lambda_0}
  \sqrt{2\pi}\times(v_{w}/c)$) in units of erg cm$^{-2}$ s$^{-1} 
  \times 10^{-13}$ for the STIS/CCD data at that wavelength ($\lambda_0$) 
  given the $S/N$ and local continuum flux ($f_\lambda$) assuming that
  the line has a Doppler width ($v_w$) of about 600 km s$^{-1}$.}  
\tablenotetext{c}{Heavy line blanketing made it difficult to sample
  the continuum so the STIS Spectroscopic Exposure Time Calculator
  (http://apt.stsci.edu/webetc/stis/stisspec.jsp) was used to compute
  the expected S/N for the spectrum.} 
\tablenotetext{d}{The detection limit is estimated for our E140M
  MAMA observation using the STIS Spectroscopic Exposure time
  calculator.  If the MAMA data were binned to the same resolution as
  the CCD data, the detection limit would be a factor of $\sqrt{2}$
  smaller.}   
\tablenotetext{e}{Predicted flux in erg cm$^{-2}$ s$^{-1} \times10^{-13}$. Calculated without considering extinction effects.}
\end{deluxetable}

\begin{deluxetable}{llcrrr}
\tablewidth{0pt}
\tabletypesize{\tiny}
\tablecolumns{6}
\tablecaption{Measured Flux and Equivalent Width of the 4680\mbox{\AA}
Feature\label{ewfluxtab}}
\tablehead{
&&\colhead{Continuum}&&\colhead{Line}&\colhead{Line EW\tablenotemark{d}}\\
\colhead{MJD}&\colhead{Year}&\colhead{Level\tablenotemark{a}}
&\colhead{$S/N$\tablenotemark{b}}
&\colhead{Flux\tablenotemark{c}}&\colhead{(\mbox{\AA})}}
\startdata
50891.4&1998.21&1.32&79&-0.12$\pm$0.22&-0.09$\pm$0.16\\
51230.5&1999.14&2.27&126&-0.47$\pm$0.23&-0.21$\pm$0.10\\
51623.8&2000.22&3.16&140&-0.10$\pm$0.30&-0.03$\pm$0.09\\
52016.8&2001.29&2.71&109&-0.63$\pm$0.32&-0.23$\pm$0.12\\
52294.0&2002.05&2.65&110&-0.26$\pm$0.31&-0.10$\pm$0.12\\
52459.5&2002.51&2.58&130&-0.29$\pm$0.26&-0.11$\pm$0.10\\
52683.1&2003.12&2.49&121&0.82$\pm$0.27&0.33$\pm$0.11\\
52727.3&2003.24&2.65&102&0.67$\pm$0.34&0.25$\pm$0.13\\
52764.3&2003.34&2.36&97&2.20$\pm$0.32&0.93$\pm$0.13\\
52778.5&2003.38&2.56&118&2.22$\pm$0.28&0.87$\pm$0.11\\
52791.7&2003.42&2.74&144&3.35$\pm$0.25&1.23$\pm$0.09\\
52813.8&2003.47&3.09&107&7.38$\pm$0.37&2.38$\pm$0.12\\
52825.4&2003.51&3.18&113&0.25$\pm$0.37&0.08$\pm$0.11\\
52852.4&2003.58&3.25&108&0.06$\pm$0.39&0.02$\pm$0.12\\
52904.3&2003.72&4.38&148&-0.03$\pm$0.34&-0.01$\pm$0.09\\
52960.6&2003.88&4.80&150&-0.03$\pm$0.37&-0.01$\pm$0.09\\
53071.2&2004.18&4.69&140&-0.28$\pm$0.44&-0.06$\pm$0.09\\
\enddata
\tablenotetext{a}{The adopted continuum level $f_\lambda$ in units of 
$10^{-12}$ erg cm$^{-2}$ s$^{-1}$ {\AA}$^{-1}$, not corrected for 
extinction.  See text.}
\tablenotetext{b}{{\it Relative\/} r.m.s.\ statistical error 
  in a 0.28 {\AA} sample of the continuum.  Estimated from dispersion 
  in the data, see Section \ref{wasitthere}. }
\tablenotetext{c}{Net flux in the 4680 {\AA} feature integrated
  from 4675 {\AA} to 4694 {\AA}, in units of 
  $10^{-12}$ erg cm$^{-1}$ s$^{-1}$.
  Positive values are emission above the continuum level.}  
\tablenotetext{d}{Measured equivalent width of 4680 {\AA} emission 
  in the range 4675--4694 {\AA}, with no assumptions
  about line shape.  Positive values are emission above the
  continuum level. }
\end{deluxetable}

\begin{deluxetable}{lcrr}
\tabletypesize{\tiny}
\tablecolumns{4}
\tablewidth{0pt}
\tablecaption{ Relative Noise Levels and Constraints on \ion{He}{2} Emission
  in STIS Data Before 2003.00 and After 2003.50\label{tableb1}}
\tablehead{ \colhead{MJD\tablenotemark{a}}  &  \colhead{Year} &
   \colhead{$S/N$\tablenotemark{b}} &
   \colhead{E.W.\ (m{\AA})\tablenotemark{c}}  }
\startdata
50891&1998.21&127&+24$\pm$36\\
51230&1999.14&133&+1$\pm$34\\
51624&2000.22&109&-50$\pm$42\\
52017&2001.29&114&-61$\pm$40\\
52294&2002.05&137&+14$\pm$34\\
52460&2002.51&119&+33$\pm$38\\
52683\tablenotemark{d}&2003.12&102&+182$\pm$45\\
52825&2003.51&137&-37$\pm$36\\
52852&2003.58&164&+44$\pm$29\\
52904&2003.72&128&+27$\pm$36\\
52961&2003.88&130&+43$\pm$35\\
53071&2004.18&156&+18$\pm$29\\
(avg)\tablenotemark{e}&\nodata&345&-11$\pm$14\\
\enddata
\tablenotetext{a}{Modified Julian Day Number}
\tablenotetext{b}{Relative r.m.s.\ noise in a 0.28 {\AA} sample, see text}
\tablenotetext{c}{Formal emission equivalent width in least-squares fit}
\tablenotetext{d}{He~II emission was clearly present at 2003.12;
  included for comparison only, see text} 
\tablenotetext{e}{Results in spectrum averaged over all these data sets
   except 2003.12, see text}
\end{deluxetable}


\clearpage
\begin{thebibliography}{}
\bibitem[Cardelli et al.(1989)]{ccm} Cardelli, J.~A., Clayton, G.~C., \& Mathis, J.~S.\ 1989, \apj, 345, 245 
\bibitem[Corcoran et al.(2001)]{xrayspec1} Corcoran, M.~F., et al.\ 2001, \apj, 562, 1031 
\bibitem[Corcoran et al.(2004)]{wings04} Corcoran, M.~F., et al.\ 2004, \apj, 613, 381 
\bibitem[Corcoran(2005)]{xraycurve} Corcoran, M.~F.\ 2005, \aj, 129, 2018 
\bibitem[Cox et al.(1995)]{cox95} Cox, P., et al.\ 1995, \aap, 297, 168 
\bibitem[de Vaucouleurs \& Eggen(1952)]{dvegg52} de Vaucouleurs, G.~\& Eggen, O.~J.\ 1952, \pasp, 64, 185 
\bibitem[Damineli(1996)]{damineli96} Damineli, A.\ 1996, \apjl, 460, L49 
\bibitem[Davidson \& Netzer(1979)]{dn79} Davidson, K., \& Netzer, H. 1979, Revs.\ Mod.\ Phys.\ 51, 715  
\bibitem[Davidson \& Humphreys(1997)]{davidsonhumphreys1997} Davidson, K.~\& Humphreys, R.~M.\ 1997, \araa, 35, 1 
\bibitem[Davidson et al.(1995)]{davidson95} Davidson, K., et al.\ 1995, \aj, 109, 1784 
\bibitem[Davidson(1999)]{kd99} Davidson, K.\ 1999, Astronomical Society of the Pacific Conference Series, 179, 304 
\bibitem[Davidson et al.(2000)]{davidson00} Davidson, K., Ishibashi, K., Gull, T.~R., Humphreys, R.~M., \& Smith, N.\ 2000, \apjl, 530, L107 
\bibitem[Davidson et al.(2001)]{kd01} Davidson, K., Smith, N., Gull, T.~R., Ishibashi, K., \& Hillier, D.~J.\ 2001, \aj, 121,1569 
\bibitem[Davidson(2002)]{kd02} Davidson, K.\ 2002, Astronomical Society of the Pacific Conference Series, 262, 267 
\bibitem[Davidson(2004a)]{etatp} Davidson, K.\ 2004a, STScI Newsletter,
  Spring 2004, 1
\bibitem[Davidson et al.(2005a)]{halpha} Davidson, K., Martin, J.~C.,
  Humphreys, R.~M., Ishibashi, K., Gull, T.~R., Stahl, O., Weis, K.,
  Hiller, D.~J., Damineli, A., Corcoran, M., \& Hamann, F.\ 2005a,
  \aj, 129, 900
\bibitem[Davidson(2005b)]{kd05c} Davidson, K.\ 2005b, Astronomical Society of the Pacific Conference Series, 332, 103
\bibitem[Fitzpatrick(1999)]{fitz} Fitzpatrick, E.~L.\ 1999, \pasp, 111, 63
\bibitem[Feast et al.(2001)]{feast01} Feast, M., Whitelock, P., \& Marang, F.\ 2001, \mnras, 322, 741 
\bibitem[Gaviola(1953)]{gaviola} Gaviola, E.\ 1953, \apj, 118, 234 
\bibitem[Gull(2005)]{gull} Gull, T.\ 2005, Astronomical Society of the Pacific Conference Series, 332, 281
\bibitem[Hamaguchi et al.(2004)]{hamaguchi04} Hamaguchi, K., Corcoran, M.~F., Gull, T., White, N.~E., Damineli, A., \& Davidson, K.\ 2004, ArXiv Astrophysics e-prints, astro-ph/0411271, to appear in ``Massive Stars in Interacting Binaries'' held in Quebec Canada (16-20 Aug, 2004)
\bibitem[Hillier et al.(2001)]{hillieretal01} Hillier, D.~J., Davidson, K., Ishibashi, K., \& Gull, T.\ 2001, \apj, 553, 837 
\bibitem[Hillier et al.(2005)]{hillier05} Hillier, D. et al.\ 2005, in preparation
\bibitem[Humphreys \& Koppelman(2005)]{rmh05} Humphreys, R.~M., \& Koppelman, M.\ 2005, Astronomical Society of the Pacific Conference Series, 332, 161 
\bibitem[Ishibashi et al.(1999)]{xrays2} Ishibashi, K., Corcoran, M.~F., Davidson, D., Swank, J.~H., Peetre, R., Drake, S.~A., Damineli, A., \& White, S.\ 1999, \apj, 525, 983
\bibitem[Ishibashi(2001)]{bish01} Ishibashi, K.\ 2001, ASP Conf.~Ser.~242: Eta Carinae and Other Mysterious Stars: The Hidden Opportunities of Emission Spectroscopy, 242, 53 
\bibitem[Johnson(1968)]{m42spect} Johnson, H.~M.\ 1968, in Stars and Stellar Systems, Chicago: University of Chicago Press, 1968, edited by Middlehurst, Barbara M.; Aller, Lawrence H., 65
\bibitem[Lamers \& Cassinelli(1999)]{windsbook} Lamers, H.~J.~G.~L.~M., \& Cassinelli, J.~P.\ 1999, Introduction to Stellar Winds / Henny J.G.L.M.~Lamers and Joseph P.~Cassinelli.~Cambridge ; New York : Cambridge University Press, 1999.~ ISBN 0521593980, p.211
\bibitem[McKenna et al.(1997)]{rrtel} McKenna, F.~C., Keenan, F.~P.,
  Hambly, N.~C., Allende Prieto, C., Rolleston, W.~R.~J., Aller,
  L.~H., \& Feibelman, W.~A.\ 1997, \apjs, 109, 225 
\bibitem[Martin(2005)]{martinconf} Martin, J.C.\ 2005, Astronomical Society of the Pacific Conference Series, 332, 114
\bibitem[Meaburn et al.(1987)]{meaburn87} Meaburn, J., Wolstencroft, R.~D., \& Walsh, J.~R.\ 1987, \aap, 181, 333 
\bibitem[Morrison \& McCammon(1983)]{xabs} Morrison, R., \& McCammon, D.\ 1983, \apj, 270, 119 
\bibitem[O'Connell(1956)]{oc56} O'Connell, D.~J.~K.\ 1956, Vistas in Astronomy, 2, 1165 
\bibitem[Osterbrock(1989)]{osterbrock} Osterbrock, D.~E.\ 1989, Astrophysics of Gaseous Nebulae and Active Galactic Nuclei, (Mill Valley, CA:  University Science Books)
\bibitem[Pittard et al.(1998)]{ecl98} Pittard, J.~M., Stevens, I.~R., Corcoran, M.~F., \& Ishibashi, K.\ 1998, \mnras, 299, L5 
\bibitem[Pittard \& Corcoran(2002)]{pc02} Pittard, J.~M.~\& Corcoran, M.~F.\ 2002, \aap, 383, 636 
\bibitem[Smith et al.(2003)]{nsmith03} Smith, N., Davidson, K., Gull, T.~R., Ishibashi, K., \& Hillier, D.~J.\ 2003, \apj, 586, 432 
\bibitem[Stahl et al.(2005)]{stahl05} Stahl, O., Weis, K., Bohmans,
  D.~J., Davidson, K., Gull, T.~R., \& Humphreys, R.~M.\ 2005, \aap,
  in press. 
\bibitem[Steiner \& Damineli(2004)]{steiner} Steiner, J.~E., \& Damineli, A.\ 2004, \apjl, 612, L133 
\bibitem[Stevens et al.(1992)]{xrays92} Stevens, I.~R., Blondin, J.~M., \& Pollock, A.~M.~T.\ 1992, \apj, 386, 265 
\bibitem[Thackeray(1953)]{thackeray} Thackeray, A.~D.\ 1953, \mnras,
  113, 211 
\bibitem[Viotti et al.(2002)]{xrayspec2} Viotti, R.~F., et al.\ 2002, \aap, 385, 874 
\bibitem[Whitelock et al.(1994)]{whitelock94} Whitelock, P.~A., Feast, M.~W., Koen, C., Roberts, G., \& Carter, B.~S.\ 1994, \mnras, 270, 364 
\bibitem[Whitelock et al.(2004)]{whitelock} Whitelock, P.~A., Feast, M.~W., Marang, F., \& Breedt, E.\ 2004, \mnras, 352, 447 
\bibitem[Zanella et al.(1984)]{zanella84} Zanella, R., Wolf, B., \& Stahl, O.\ 1984, \aap, 137, 79 
\bibitem[Zethson(2001)]{zethsonphd} Zethson, T.\ 2001, Ph.D.~Thesis, Lunds Universitet
\end{thebibliography}
\end{document}